\documentclass[Journal]{IEEEtranTIE}

\usepackage{graphicx}          
\usepackage{graphicx}
\usepackage{subfigure}
\usepackage{cite}
\usepackage{picinpar}
\usepackage{amsmath}

\usepackage{amssymb}
\usepackage{algpseudocode}
\usepackage{url}
\usepackage{flushend}
\usepackage[latin1]{inputenc}
\usepackage{colortbl}
\usepackage{soul}
\usepackage{multirow}
\usepackage{pifont}
\usepackage{color}
\usepackage{alltt}
\usepackage[hidelinks]{hyperref}
\usepackage{enumerate}
\usepackage{siunitx}
\usepackage{breakurl}
\usepackage{epstopdf}
\usepackage{pbox}

\usepackage{lipsum}
\usepackage{cite}
\usepackage{tikz}
\usepackage{caption}
\usepackage{multirow}
\usepackage{tabulary}
\usepackage{tabularx}
\usepackage{siunitx}
\usepackage{pifont}
\usepackage{makecell}

\usepackage{amsfonts}
\usepackage{stackengine}
\def\delequal{\mathrel{\ensurestackMath{\stackon[1pt]{=}{\scriptscriptstyle\Delta}}}}
\newtheorem{theorem}{Theorem}
\newtheorem{assumption}{Assumption}
\newtheorem{remark}{Remark}

\newtheorem{lemma}{Lemma}
\newtheorem{definition}{Definition}

\usepackage{titlesec}
\titlespacing*{\section} {0pt}{1ex}{1ex}
\titlespacing*{\subsection} {0pt}{1ex}{1ex}
\titlespacing*{\subsubsection} {0pt}{1ex}{1ex}

\begin{document}

	\title{Distributed Optimal Coverage Control in Multi-agent Systems: Known and Unknown Environments} 
	\author{
		\vskip 1em
		
		\author{Mohammadhasan~Faghihi,  Meysam~Yadegar, Mohammadhosein~Bakhtiaridoust, Nader~Meskin, Javad~Sharifi and Peng~Shi
			
			\thanks{M. Faghihi, M. Yadegar,   M. Bakhtiaridoustare and J. Sharifi with the Department
				of Electrical and Computer Engineering, Qom University of Technology, Qom,
				Iran, e-mail: faghihi.mh@qut.ac.ir; yadegar@qut.ac.ir: bakhtiaridoust.mh@qut.ac.ir; sharifi@qut.ac.ir }
			\thanks{N. Meskin is with the Department of Electrical Engineering, Qatar University, Doha, Qatar, e-mail: nader.meskin@qu.edu.qa.}
			\thanks{P. Shi is with the School of Electrical and Electronic Engineering, University of Adelaide, Adelaide, SA 5005, Australia, e-mail: peng.shi@adelaide.edu.au.}
		}
	}

	\maketitle
	\begin{abstract}                          
		This paper introduces a novel approach to solve the coverage optimization problem in multi-agent systems. The proposed technique \textcolor{black}{  offers an optimal solution  with a lower cost with respect to conventional Voronoi-based techniques by effectively handling the issue of agents remaining stationary in regions void of information using a ranking function.} The proposed approach leverages a novel cost function for optimizing the agents coverage and the cost function eventually  aligns with the conventional Voronoi-based cost function. Theoretical analyses are conducted to assure the asymptotic convergence of agents towards the optimal configuration. 
		A distinguishing feature of this approach lies in its departure from the reliance on geometric methods that are characteristic of Voronoi-based approaches; hence can be implemented more simply. Remarkably, the technique is adaptive and applicable to various environments with both known and unknown information distributions. Lastly, the efficacy of the proposed method is demonstrated through simulations, and the obtained results are compared with those of Voronoi-based algorithms.
	\end{abstract}

	\section{Introduction}
	A multi-agent system is a collection of autonomous agents that interact with each other and their environment to achieve individual and/or collective goals, and such systems have found applications across various fields \cite{yadegar2021fault,yadegar2021output,yadegar2021mission}.
	
	One of the fundamental topics in multi-agent systems is the coverage control problem \cite{kantaros2015distributed,B_li2005distributed,C_hokayem2007persistent,song2020coverage}. Coverage control involves distributing agents in such a way that they collectively explore and monitor a given environment efficiently and effectively. In other words, the objective is to position the agents within an environment in such a way that they can gather the maximum amount of data sources or information. Additionally, this positioning strategy aims to prevent overlap between agents' coverage areas. In order to formulate the coverage control problem,  the information in the environment is modeled by a mathematical function called the \textit{density function} \cite{global, D_leonard2013nonuniform, G_caicedo2008coverage}, and the desired agents configuration metric  is modeled by minimizing a function named the \textit{cost function}.
	The coverage environment  can either be known or unknown in terms of agents' knowledge of the density function. In a known environment, the density function of each point in the environment is known for every agent, and the coverage control problem involves arranging the positions of agents to optimally sense regions of interest based on some specific cost functions. However, in an unknown environment, the density function is not priorly known for each agent, and agents, therefore, must learn the density function of the environment by identifying interesting regions and gathering information about them during their movement.
	
	In general, two primary approaches are employed to address the coverage problem: the greedy-evolutionary and Voronoi-based methods \cite{cassandras1}. 
	The former combine elements of both greedy algorithms, which make locally optimal choices at each step, and evolutionary algorithms, which mimic natural selection processes to find better solutions over time. They aim to iteratively improve coverage by selecting and evolving agent positions.
	In \cite{cassandras1}, a greedy algorithm is employed with two performance bounds using partial and greedy curvature to construct the coverage component of the objective function. In \cite{pso1}, a novel particle swarm optimization (PSO) technique is used to address the sensor deployment problem, and in \cite{ga2}, a genetic algorithm is developed for the coverage path planning of unmanned aerial vehicles (UAVs) while minimizing their energy consumption.
	Although these approaches may yield \textcolor{black}{nearly optimal} solutions, the computational cost is prohibitively high, making them inappropriate for online and real-time applications.
	
	On the other hand, Voronoi-based methods \cite{bullo4,sch4,EB1},
	make a trade-off between potential optimality and efficiency, opting to prioritize efficiency over achieving the absolute best solution. This approach is particularly valuable in scenarios where the real-time performance or the scalability of the solution is crucial.
	
	A Lloyd approach based on Voronoi partition was first adapted in \cite{bullo4} for the optimal coverage problem in autonomous vehicle networks performing distributed sensing tasks.
	In \cite{est1},  the Gaussian process upper confidence bound (GP-UCB) algorithm is combined with the Lloyd control algorithm to achieve coverage in an unknown domain.
	In \cite{sch4}, a controller is proposed  that utilizes parameter adaptation to address coverage tasks in unknown sensory environments, while also incorporating a consensus term within the parameter adaptation law to propagate sensory information among the robots in the network.
	\textcolor{black}{The research conducted in \cite{sch12} suggests an online adaptive distributed controller that relies on implementing a Voronoi-based cost function using gradient descent for coverage tasks.}
	The work in \cite{iet1} introduces an adaptive spatial estimation algorithm for the mobile sensor network with the main aim of estimating the density function and guiding  the sensors to their optimal positions.
	The work described in \cite{sch10} proposes a distributed motion control algorithm in which the robots proceed asynchronously with dead-zone-based robustification.
	{The work in \cite{E_todescato2017multi} proposes a strategy that smoothly transitions from estimation to coverage, resulting in improved transient behavior compared to traditional approaches. It leverages the superior estimation performance of non-parametric Gaussian regression over parametric methods, while also bounding its computational complexity.}
	A time-varying density function was used to describe the time cost metric of a robot network with an input disturbance in \cite{ft} for a dynamic environment.
	In \cite{EB1},  a Bayesian prediction-based coverage control strategy is presented for a multi-agent system where it takes into account noise and is capable of approximating time-varying density functions.
	In \cite{DR1}, the informative coverage algorithm is proposed by considering the non-holonomic limitations and the unknown dynamics of the mobile robots.
	A coverage method for unicycle robots is also suggested in \cite{oc}, while taking outside disturbance into account.
	In \cite{BH3}, taking into account the input and state constraints, a method based on the combination of predictive model control and Voronoi coverage control is presented.
	{The work in \cite{F_rickenbach2024active} propose a tracking MPC-based coverage hierarchical framework, which includes safety-ensuring constraints, and avoids the difficult design of terminal constraints needed,}
	In \cite{roi3}, a novel team-based strategy is presented that allows the formation of several teams of agents within the coverage control framework with the heterogeneity in their embedded communication capabilities or dynamics. A distributed, self-triggered control strategy for centroidal Voronoi coverage control is presented in \cite{selft}  where the agent sampling or communication instants are reduced without affecting the performance of the mobile sensor network.

	While the efficiency of this approach proves beneficial for scenarios demanding real-time performance, it does encounter an issue with the localized optimality of the solution. Moreover, if an agent is situated within a Voronoi partition that lacks any sensory information, the agent will remain stationary indefinitely. This circumstance results in inefficient resource allocation for environmental coverage.

	In this paper, based on the above discussion, we introduce an innovative distributed coverage technique which aims at resolving the challenges outlined earlier.  {Our approach reaches a lower cost than that of the Voronoi-based methods by circumventing the issue of agents remaining stationary when surrounded by information voids. Specifically, by incorporating the effect of all points of the environment in a fuzzy fashion, instead of confining each agent to its Voronoi partition, we ensure that agents don't get trapped when they encounter areas with no information.}
	Moreover, we reframe the problem by formulating the optimization of agent composition using a novel metric. This metric progressively converges to the conventional Voronoi-based cost function as time extends towards infinity. In our analysis, we demonstrate the asymptotic convergence of agents towards their optimal configuration. Notably, our method is versatile, applicable to both known and {unknown environments featuring areas with no information.}
	Furthermore, unlike Voronoi-based methods, our presented approach circumvents the need for geometric techniques, such as determining Voronoi diagram partitioning edges. As a result, our method can be implemented with greater simplicity.

	The rest of the paper is organized as follows: The notations and preliminaries are described in Section \ref{s_2}. Section \ref{s_3}  addresses the   problem formulation. Section \ref{s_4} presents the proposed coverage control algorithm for both known and unknown environments. 
	In Section \ref{s_5}, using different simulation scenarios, the   performance of the proposed method is evaluated, and a  comparison with the Voronoi-based method is conducted.
	Finally, Section \ref{s_6} concludes the paper.
	
	\section{Notations and Preliminaries}\label{s_2}
	In order to show the problem addressed in this paper, the conventional representation of required variables such as matrices and vectors is presented in bold. $\mathbb{R}$ and $\mathbb{R}_{\ge 0}$, respectively, stand for the real number field and the set of non-negative real numbers. The set of $n \times 1$ real column vectors is denoted by $\mathbb{R}^n$ and 
	the set of $n \times m$ real matrices is denoted by $\mathbb{R}^{n \times m}$. Furthermore,  $\mathbf{I}_n$ represents $n \times n$ identity matrix, $\mathbf{0}_{n \times m}$ represents $n \times m$ zero matrix, $\mathbf{1}_{n \times m}$ represents $n \times m$ one matrix,  $(\cdot)^{\mathrm{T}}$ represents transpose, $|\cdot|$ is the absolute value, and $\lVert \cdot \rVert$ is the Euclidean norm. 
	
	The directed graph $\mathcal{G}=(\mathcal{V},\mathcal{E})$, where $\mathcal{V}=\{v_1,v_2,\dots,v_n\}$ is the collection of agent indexes and $\mathcal{E} \in \mathcal{V} \times \mathcal{V}$ is the set of edges connecting different agents, serves as a representation of the communication topology among $n$ agents. If $(v_i,v_j)\in \mathcal{E}$, it means that the agent $j$ and the agent $i$ are neighbors and that the agent $j$ can obtain information from the agent $i$. The presumption is that self edges, i.e., $(v_i,v_i)$, is not permitted. Suppose, a function $l: \mathcal{V}\times \mathcal{V} \to \mathbb{R}$ such that
	$l_{i,j}>0$ if $(v_i,v_j)\in \mathcal{E}$, and $l_{i,j}=0$ if $(v_i,v_j)\notin \mathcal{E}$, where $l_{i,j}\ge 0$ represents the strength of communication between agents $i$ and $j$.
	
	Additionally, $\mathbf{L} \in \mathbb{R}^{n \times n}$ is the system's weighted graph Laplacian  and is defined as   
	\begin{equation}
		\mathbf{L}_{i,j}=
		\begin{cases}
			-l_{i,j}, & \text{for} \quad i \neq j, \\
			\sum_{j=1}^{n}l_{i,j}, & \text{for} \quad i = j. \\
		\end{cases}
	\end{equation}
	If the agents network is connected, the graph's Laplacian is positive semi-definite, and it has precisely one zero eigenvalue with the associated  eigenvector of 
	$\mathbf{1}_{n \times 1}$.
	Furthermore, $\kappa(\cdot)$ is the Heaviside function and is defined as follows:
	\begin{equation}
		\label{fo18}
		\kappa(c)=
		\begin{cases}
			1, & \text{for} \quad c>0 , \\
			0, & \text{for} \quad c \le 0.
		\end{cases}
	\end{equation}
	In addition, $\delta(\cdot)$ is the derivative of $\kappa(\cdot)$, and $	\delta(x)=0, \quad  \forall x \neq 0,$ and $\int \delta(x)\mathrm{d}x=1$.
	\section{Problem Formulation}\label{s_3}	
	In the following, it is assumed that there exist $n$ agents in a  bounded and convex environment denoted by $\mathbf{Q}\subseteq \mathbb{R}^N$.
	An arbitrary point in $\mathbf{Q}$ is denoted by $\mathbf{q}$.
	Furthermore, the position of each agent in the environment at time $t$ is represented by $\mathbf{p}_i(t)\in \mathbf{Q}$, $i = 1,\dots,n$.
	\begin{definition}
		\textbf{(Voronoi partition)}
		The Voronoi partition of the environment  $\mathbf{Q}$ generated by the $i$-th agent at time $t$ named as the generator point of Voronoi is defined as \cite{bullo4}:
		\begin{align}
			V_i(t)\!=\!\{\mathbf{q} \in \mathbf{Q} \mid \lVert \mathbf{q}-\mathbf{p}_i(t) \rVert &\le \lVert \mathbf{q}-\mathbf{p}_j(t) \rVert,\\\nonumber& j=1,\dots,n,j \neq i\}.
			\label{fo4}
		\end{align}
	\end{definition}
	{It is assumed that each agent is able to compute its Voronoi partition. For this purpose, each agent need to know its location and other agents locations \cite{bullo4}.}
	\begin{definition}
		\textbf{(Density function)}
		The density function is defined as $\phi: \mathbf{Q} \to {\mathbb{R}}_{\ge 0}$ with the following property:
		\begin{equation}
			0<\int_{\mathbf{Q}}\phi(\mathbf{q})\mathrm{d}\mathbf{q} \le \eta<\infty,
			\label{fo4_2}
		\end{equation} 	
		where the larger values of $\phi(\mathbf{q})$ correspond to more important areas. 
	\end{definition}
	\begin{definition}
		\textbf{(Known environment)}
		A known environment is an environment where the agent knows the value of $\phi(\mathbf{q})$ for each $\mathbf{q} \in \mathbf{Q}$. In fact, the agent does not require to measure information using the relevant sensors.
	\end{definition}
	\begin{definition}
		\textbf{(Unknown environment)}
		The unknown environment is the environment where
		the value of  $\phi(\mathbf{q})$, $\forall  \mathbf{q} \in \mathbf{Q}$, is not priorly  known for each agent, and consequently each agent needs to  measure or estimate it by its relevant sensors.
	\end{definition}
	\begin{definition}
		\textbf{(Zero information area)}
		A subset $\Omega \subset Q$ is called a zero information area if  $\phi(q)=0, \forall q \in \Omega$.
	\end{definition}
	\begin{definition}
		\textbf{(Separated information area)}
		\label{def:separate_area}
		$\mathbf{Q}$ is called an environment with separated information areas if it can be partitioned into  compact  sets $\mathbf{A}_l$, $l=1,2,\dots,l_n$, and contains at least one zero information area.
	\end{definition}
	Conventionally, the converge problem is modeled based on the Voronoi diagram and the following cost function \cite{bullo4,sch4}:
	\begin{equation}
		\label{f1}
		H\big(\mathbf{P}(t)\big)\triangleq\frac{1}{2}\sum_{i=1}^{n}\int_{V_i(t)}\lVert \mathbf{q}-\mathbf{p}_i(t) \rVert^2\phi(\mathbf{q})\mathrm{d}\mathbf{q},
	\end{equation}
	such that $	H\big(\mathbf{P}(t)\big) >0, \quad t\geq 0$, 
	where  $\mathbf{P}(t)\triangleq \{\mathbf{p}_1(t),\dots,$
	$\mathbf{p}_n(t)\}$.
	{\begin{definition}\textbf{(Optimal coverage configuration)}
			An agent network is said to be in the (locally) optimal
			coverage configuration if every agent is positioned in the environment such that the cost function \eqref{f1} is (locally) minimized.
	\end{definition}}
	Assume the agents' locations obey a first-order dynamical behavior described by 	
	\begin{equation}
		\label{f15_2}
		\dot{\mathbf{p}}_{i}(t)=\mathbf{u}_{i}(t), \quad i=1,\dots, n,
	\end{equation}
	where $\mathbf{u}_{i}(t) \in  \mathbb{R}^N$ is the control input.
	In order to find the minimum value of cost function for calculating the position  of each agent, the gradient descent method is used as follows \cite{bullo4,sch14}:
\begin{align}
	\dot{\mathbf{p}}_i(t)
	&=-k_i	\frac{\partial H\big(\mathbf{P}(t)\big)}{\partial \mathbf{p}_{i}(t)}  \notag
	= k_i\int_{V_i(t)} (\mathbf{q}-\mathbf{p}_{i}(t))\phi(\mathbf{q}) \mathrm{d}\mathbf{q} \notag	\\
	&=k_{\mathrm{prop}_i}(t) \big(\mathbf{C}_{V_i}(t)-\mathbf{p}_i(t) \big),
	\label{f1_45}
\end{align}
	where $k_i$ and $k_{\mathrm{prop}_i}(t) \triangleq k_i\int_{V_i(t)}\phi(\mathbf{q})\mathrm{d}\mathbf{q}$  are  positive   gains and $\mathbf{C}_{V_i}(t)$ is the centroid of a Voronoi partition obtained as follows:
	\begin{equation}
		\mathbf{C}_{V_i}(t)=\frac{\int_{V_i(t)}\mathbf{q}\phi(\mathbf{q})\mathrm{d}\mathbf{q}}{\int_{V_i(t)}\phi(\mathbf{q})\mathrm{d}\mathbf{q}}.
		\label{c1}
	\end{equation}
	Equation \eqref{f1_45} implies that each agent is \textcolor{black}{moving towards} the centroid of its Voronoi partition. 
	
	One of the major drawbacks of utilizing the Voronoi-based methods given by \eqref{f1_45} and \eqref{c1} to position agents appropriately is 
	when dealing with large environment that contain zero information areas as Voronoi partitions. In this case, the control input of the relevant agent can not be computed using \eqref{c1}.
	This problem is rooted in the fact that the agents are confined to their own Voronoi partition and cannot leverage the available information in other Voronoi partitions.
	To circumvent this problem, first the function \eqref{f1} is reformulated based on the whole environment as follows:
	\begin{equation}
		\label{f1_2}
		H\big(\mathbf{P}(t)\big)=\frac{1}{2}\sum_{i=1}^{n}\int_{\mathbf{Q}}h_i\big(\mathbf{q},\mathbf{P}(t)\big) \lVert \mathbf{q}-\mathbf{p}_{i}(t) \rVert^2\phi(\mathbf{q})\mathrm{d}\mathbf{q},
	\end{equation}
	where
	\begin{equation}
		\label{f1_3}
		{h}_i\big(\mathbf{q},\mathbf{P}(t)\big)=
		\begin{cases}
			1 , & \text{if} \quad \lVert \mathbf{q}-\mathbf{p}_{i}(t) \rVert \le \lVert \mathbf{q}-\mathbf{p}_{j}(t) \rVert,\quad\forall j, \\
			0 , & \text{otherwise}.
		\end{cases}
	\end{equation}
	Comparing \eqref{f1} and 	\eqref{f1_2}, the Voronoi partition is modeled by $h_i(\mathbf{q},\mathbf{P}(t)), i=1,\dots,n$. {Using (\ref{f1_2}), each agent utilizes the information from its Voronoi partition to compute its control input, as previously mentioned. However, our objective is to develop an alternative cost function capable of leveraging comprehensive environmental information to address the aforementioned issues. It is important to highlight that in both the Voronoi-based method and our proposed framework, every agent must have knowledge of the positions of all agents. Furthermore, in scenarios where the density function is unknown, our method performs distributed estimation of environmental information and subsequently optimizes the cost function to determine the control input.}
	\section{Proposed Method }\label{s_4}	
	In this section, the proposed algorithm is presented for two cases, namely known and unknown  environments. For this purpose, first, some definitions and lemmas are needed to be introduced.
	\begin{definition} \label{definition:4}
		\textbf{(Ranking function)}
		\textit{The ranking function of the $i$-th agent $R_i: \mathbf{Q}\times \mathbf{P} \to\{0,\dots,n-1\}$ is defined as follows:}
		\begin{align} 
			\label{f3}
			R_i\big(\mathbf{q},\mathbf{P}(t)\big)\!\delequal\!
			\sum_{l=1}^{n}\kappa(\lVert \mathbf{q}-\mathbf{p}_{i}(t) \rVert^2-\lVert \mathbf{q}-\mathbf{p}_{l}(t) \rVert^2),\\\nonumber i=1,\dots,n.
		\end{align}
	\end{definition}

	Based on  the definition of $R_i\big(\mathbf{q},\mathbf{P}(t)\big)$,  each agent $i$ at time $t\geq 0$ assigns a rank  $R_i\big(\mathbf{q},\mathbf{P}(t)\big)$ between  $0$ to $n-1$ to all the points in the environment, where the points that the closest agent is the $i$-th agent receive $(0)$-rank, i.e. $R_i\big(\mathbf{q},\mathbf{P}(t)\big)=0$, the points that the second closest agent is the $i$-th agent receive $(1)$-rank, i.e.  $R_i\big(\mathbf{q},\mathbf{P}(t)\big)=1$, and the points that the farthest agent is the $i$-th agent receive at most $(n-1)$-rank, i.e.  $R_i\big(\mathbf{q},\mathbf{P}(t)\big)=n-1$. {It should be noted that points with the same distance from two agents, i.e., $\lVert \mathbf{q}-\mathbf{p}_{j}(t) \rVert = \lVert \mathbf{q}-\mathbf{p}_{i}(t) \rVert$, are assigned the same rank by each of those agents, which translates to $R_i=R_j$ for that particular point. These points lie on the perpendicular bisector of $\mathbf{p}_{i}$ and $\mathbf{p}_{j}$ in $\mathbf{Q}$.}
	In this work, the following cost function is introduced:
	\begin{align}
		\label{f2}
		\hat{H}\big(t,\mathbf{P}(t)\big)
		\!\triangleq\!\frac{1}{2}\sum_{i=1}^{n}\int_{\mathbf{Q}}h_{\lambda}\big(t,R_i(\mathbf{q},\mathbf{P}(t))\big)
		\lVert \mathbf{q}-\mathbf{p}_{i}(t) \rVert^2\phi(\mathbf{q})\mathrm{d}\mathbf{q},
	\end{align}
	where
	\begin{equation}
		h_{\lambda}\big(t,R_i(\mathbf{q},\mathbf{P}(t))\big)=\text{exp}\Big\{\frac{-R_i\big(\mathbf{q},\mathbf{P}(t)\big)}{\lambda(t)}\Big\}	
		\label{f2_2},
	\end{equation}
	and {$\lambda(t):\mathbb{R}_{\ge 0}\to \mathbb{R}_{\ge 0}$ where $\lim_{t\to \infty}\lambda(t)=0$}. {Note that the function \(\lambda: [0, \infty) \to \mathbb{R}\) is continuously differentiable, strictly positive, strictly decreasing to zero ($\dot{\lambda}(t)<0, t\geq 0$) with a bounded second derivative, and satisfies \(\lambda(0) = \lambda_0 > 0\), where \(\lambda_0\) is finite.}
	{
		\begin{lemma}\label{LemmaB}
			Given the definition of \eqref{f2_2}, $\forall i,j\in\{1,\cdots,n\}$, the following equation holds  
			\begin{align}\label{bt}
				\frac{1}{2}{\sum_{j=1}^{n}}\int_{\mathbf{Q}}
				\frac{\partial h_{\lambda}\!(t, R_j)}{\partial \mathbf{p}_{i}(t)}
				\lVert \mathbf{q}-\mathbf{p}_{j}(t)\rVert^2 \phi(\mathbf{q})\mathrm{d}\mathbf{q} = 0.
			\end{align}
			\end{lemma}
		
			\textbf{Proof.}
			Using (\ref{f3}) and chain rule, left-side of \eqref{bt} can be written as
		\begin{align}
	\nonumber 
	\label{expB}
	\frac{1}{2}&\sum_{j=1}^{n}\int_{\mathbf{Q}}\bigg[\frac{\partial h_{\lambda}\!(t, R_j)}{\partial R_j}
	\lVert \mathbf{q}-\mathbf{p}_{j}(t) \rVert^2\phi(\mathbf{q})\\
	\nonumber&\sum_{l=1}^{n}\!\frac{\partial \big(\lVert \mathbf{q}\!-\!\mathbf{p}_{j}(t)\rVert^2\big)}{\partial \mathbf{p}_{i}(t)}\delta\big(\lVert \mathbf{q}\!-\!\mathbf{p}_{j}(t) \rVert^2\!\!\!-\!\lVert \mathbf{q}\!-\!\mathbf{p}_{l}(t) \rVert^2\big)\!\bigg]\mathrm{d}\mathbf{q}\\
	-\frac{1}{2}&\sum_{j=1}^{n}\int_{\mathbf{Q}}\bigg[\frac{\partial h_{\lambda}\!(t, R_j)}{\partial R_j}
	\lVert \mathbf{q}-\mathbf{p}_{j}(t) \rVert^2\phi(\mathbf{q})\\
	\nonumber&\sum_{l=1}^{n}\!\frac{\partial\big(\lVert \mathbf{q}\!-\!\mathbf{p}_{l}(t) \rVert^2\big)}{\partial \mathbf{p}_{i}(t)}\delta\big(\lVert \mathbf{q}\!-\!\mathbf{p}_{j}(t) \rVert^2\!\!\!-\!\lVert \mathbf{q}\!-\!\mathbf{p}_{l}(t) \rVert^2\big)\!\bigg]\mathrm{d}\mathbf{q}.
\end{align}
The terms $\frac{\partial \big(\lVert \mathbf{q}-\mathbf{p}_{j}(t)\rVert^2\big)}{\partial \mathbf{p}_{i}(t)}$ and $\frac{\partial \big(\lVert \mathbf{q}-\mathbf{p}_{l}(t)\rVert^2\big)}{\partial \mathbf{p}_{i}(t)}$ in (\ref{expB}) are non-zero if $i=j$ and $i=l$, respectively. Thus, we have
\begin{align}
	\nonumber
	\label{expB2}
	-&\int_{\mathbf{Q}}\bigg[\frac{\partial h_{\lambda}\!(t, R_i)}{\partial R_i}
	\lVert \mathbf{q}-\mathbf{p}_{i}(t) \rVert^2\phi(\mathbf{q})\big(\mathbf{q}-\mathbf{p}_{i}(t)\big)\\
	\nonumber&\sum_{l=1}^{n}\delta\big(\lVert \mathbf{q}-\mathbf{p}_{i}(t) \rVert^2-\lVert \mathbf{q}-\mathbf{p}_{l}(t) \rVert^2\big)\bigg]\mathrm{d}\mathbf{q}\\
	+&\sum_{j=1}^{n}\int_{\mathbf{Q}}\bigg[\frac{\partial h_{\lambda}\!(t, R_j)}{\partial R_j}
	\lVert \mathbf{q}-\mathbf{p}_{j}(t) \rVert^2\phi(\mathbf{q})\\
	\nonumber&\big(\mathbf{q}-\mathbf{p}_{i}(t)\big)\delta\big(\lVert \mathbf{q}-\mathbf{p}_{j}(t) \rVert^2-\lVert \mathbf{q}-\mathbf{p}_{i}(t) \rVert^2\big)\bigg]\mathrm{d}\mathbf{q}.
\end{align}
The second term in (\ref{expB2}) is non-zero if $\lVert \mathbf{q}-\mathbf{p}_{j}(t) \rVert = \lVert \mathbf{q}-\mathbf{p}_{i}(t) \rVert$, which translates to $R_i=R_j$. Therefore, we can write
\begin{align}
	\nonumber
	\label{expB3}
	-&\int_{\mathbf{Q}}\bigg[\frac{\partial h_{\lambda}\!(t, R_i)}{\partial R_i}
	\lVert \mathbf{q}-\mathbf{p}_{i}(t) \rVert^2\phi(\mathbf{q})\big(\mathbf{q}-\mathbf{p}_{i}(t)\big)\\
	\nonumber&\sum_{l=1}^{n}\delta\big(\lVert \mathbf{q}-\mathbf{p}_{i}(t) \rVert^2-\lVert \mathbf{q}-\mathbf{p}_{l}(t) \rVert^2\big)\bigg]\mathrm{d}\mathbf{q}\\
	\nonumber+&\int_{\mathbf{Q}}\bigg[\frac{\partial h_{\lambda}\!(t, R_i)}{\partial R_i}
	\lVert \mathbf{q}-\mathbf{p}_{i}(t) \rVert^2\phi(\mathbf{q})\big(\mathbf{q}-\mathbf{p}_{i}(t)\big)\\
	&\sum_{j=1}^{n}\delta\big(\lVert \mathbf{q}-\mathbf{p}_{j}(t) \rVert^2-\lVert \mathbf{q}-\mathbf{p}_{i}(t) \rVert^2\big)\bigg]\mathrm{d}\mathbf{q}.
\end{align}
Finally, since $\delta(x)=\delta(-x)$, the first and second terms cancel out, and the proof is completed.\hfill$\blacksquare$			
	}
	{
		\begin{remark}\label{remark:lambda}
			\textcolor{black}{The choice of the parameter $\lambda(t)$ is arbitrary.} However, exponential functions, such as $\lambda(t)=\lambda_{\mathrm{s}}\alpha^{-\lambda_{\mathrm{f}}t}$ where $\lambda_\mathrm{s},\lambda_\mathrm{f}>0$, and $\alpha>1$, are often considered due to their mathematical simplicity and their ability to model various types of dynamics.
	\end{remark}}
	\begin{remark}\label{remark:2}
		Unlike 	$h_i(\cdot)$ given by \eqref{f1_3} which has a binary value,  $h_{\lambda}(\cdot)$ can have any value in the range of $(0,1]$. According to Definition \ref{definition:4},
		the	closest agents to the point $\mathbf{q} \in \mathbf{Q}$ 
		gets the maximum value of $h_{\lambda}(\cdot)$, while the remaining agents receive $0<h_{\lambda}<1$. 
	\end{remark}
	\begingroup
	\color{black}
	\begin{lemma} \label{lemma:2}
		\textit{Considering \eqref{f1_2} and \eqref{f2}, for any  configuration of agents $\mathbf{P}\triangleq \{\mathbf{p}_1,\dots,\mathbf{p}_n\}$, the following equation holds:} 
		\begin{equation}\label{41}
			\lim_{t\to \infty}\hat{H}\big(t,\mathbf{P}\big)=H\big(\mathbf{P}\big).
		\end{equation}
	\end{lemma}
	
	\textbf{Proof.}
	According to the definition of $\lambda(t)$ given by \eqref{f2_2}, we can deduce that $\lambda(t) \to 0$	as $t \to \infty$.  Moreover, based on Definition \ref{definition:4}, $R_i(\mathbf{q},\mathbf{P})=0$, if $\lVert \mathbf{q}-\mathbf{p}_{i} \rVert \le \lVert \mathbf{q}-\mathbf{p}_{j} \rVert$, $\forall j$.  Hence,
	\begin{align}
		\lim_{t\to \infty}&h_{\lambda}\big(t, R_i(\mathbf{q},\mathbf{P})\big)=
		\begin{cases}
			1 ,\!\!\!\! & \text{if} \  \lVert \mathbf{q}-\mathbf{p}_{i} \rVert \le \lVert \mathbf{q}-\mathbf{p}_{j} \rVert,~\forall j, \\
			0 ,\!\!\!\! & \text{otherwise},
		\end{cases}    
		\label{L3}
	\end{align}
	therefore, using \eqref{f1_3}, we have $	\lim_{t\to \infty}h_{\lambda}\!\big(t, R_i(\mathbf{q},\mathbf{P})\big)={h}_i(\mathbf{q},\mathbf{P}),$ which leads to \eqref{41}.\hfill$\blacksquare$
	\endgroup
	
	For ease of notation, we denote $h_{\lambda}\!(t, R_i)$ and $\hat{H}(t)$ for $ h_{\lambda}\!\big(t, R_i(\mathbf{q},\mathbf{P}(t))\big)$ and $\hat{H}\big(t,\mathbf{P}(t)\big)$, respectively.
	\subsection{ Known Environment}
	In this section, a known environment is considered and a control signal for each agent is designed such that the proposed cost function $\hat{H}(\cdot)$ introduced by 	\eqref{f2} is minimized. For this purpose, similar to the Voronoi-based algorithms, the gradient descent method is used to update the position of each agent as follows:
	\begin{equation}
		\label{f15}
		\mathbf{u}_{i}(t)=-\epsilon\frac{\partial \hat{H}(t)}{\partial \mathbf{p}_{i}(t)}, \quad i=1,\dots, n,
	\end{equation}
	where $\epsilon$ is a positive scalar gain.
	The following lemma shows the calculation of the right-hand side of \eqref{f15}.
	
	\begin{lemma}\label{lemma:3}
		\textit{Consider the proposed cost function given by \eqref{f2}. The following equation holds:}
		\begin{equation}
			\label{fo15}
			\frac{\partial \hat{H}(t)}{\partial \mathbf{p}_{i}(t)}=-\int_{\mathbf{Q}} h_{\lambda}\!(t, R_i)(\mathbf{q}-\mathbf{p}_{i}(t))\phi(\mathbf{q}) \mathrm{d}\mathbf{q}.
		\end{equation}
	\end{lemma}
	\textbf{Proof.}
	According to \eqref{f2}, we can write
\begin{align}\nonumber
	\frac{\partial \hat{H}(t)}{\partial \mathbf{p}_{i}(t)}=&\frac{1}{2}\sum_{j=1}^{n}\int_{\mathbf{Q}} \bigg[h_{\lambda}\!(t,\textcolor{black}{ R_j})
	\frac{\partial (\lVert \mathbf{q}-\mathbf{p}_{j}(t) \rVert^2) }{\partial \mathbf{p}_{i}(t)}
	\\
	&\qquad+\frac{\partial h_{\lambda}\!(t, \textcolor{black}{ R_j})}{\partial \mathbf{p}_{i}(t)}\lVert \mathbf{q}-\mathbf{p}_{j}(t)\rVert^2\Bigg]
	\phi(\mathbf{q})\mathrm{d}\mathbf{q}.
	\label{p2}
\end{align}	
Using the chain rule and some manipulations, the following is obtained.
\begin{align}
	&\frac{\partial \hat{H}(t)}{\partial \mathbf{p}_{i}(t)}\!=\!-\!\int_{\mathbf{Q}}h_{\lambda}\!(t, R_i)(\mathbf{q}-\mathbf{p}_{i}(t))\phi(\mathbf{q})\mathrm{d}\mathbf{q}
	\notag
	\\
	&\quad\qquad\quad+	\frac{1}{2}{\sum_{j=1}^{n}}\int_{\mathbf{Q}}
	\frac{\partial h_{\lambda}\!(t, R_j)}{\partial \mathbf{p}_{i}(t)}
	\lVert \mathbf{q}-\textcolor{black}{\mathbf{p}_{j}(t)}\rVert^2 \phi(\mathbf{q})\mathrm{d}\mathbf{q}.
	\label{p3}
\end{align}
	{Based on Lemma \ref{LemmaB}, the second term of the right-hand side of \eqref{p3} is vanished and the proof is completed. \hfill $\blacksquare$	}
	
	Using Lemma \ref{lemma:3} and \eqref{fo15}, the control law of the proposed method is obtained as follows:	
	\textcolor{black}{
	\begin{equation}
		\label{f16}
		\mathbf{u}_{i}(t)=\epsilon \mathbf{W}_i(t,\mathbf{P}(t)),
	\end{equation}}
	where
	\textcolor{black}{
	\begin{equation} \label{eq:32}
		\mathbf{W}_{i}(t,\mathbf{P}(t)) \triangleq \int_{\mathbf{Q}} h_{\lambda}\!(t, R_i)\big(\mathbf{q}-\mathbf{p}_{i}(t)\big)\phi(\mathbf{q}) \mathrm{d}\mathbf{q}.
	\end{equation}}

\begingroup
\color{black} 
The following lemma gives some facts about $\mathbf{W}_{i}(t,\mathbf{P}(t))$.
\begin{lemma}\label{lemma:4}
	 $\mathbf{W}_i(t,\mathbf{P}(t))$ defined in \eqref{eq:32} and  $\frac{\partial \mathbf{W}_i(t)}{\partial \mathbf{p}_j(t)}$ are bounded for all $i,j \in \{1,\ldots,n\}$ and $t \geq 0$.
\end{lemma}

\textbf{Proof.}
 Since $\mathbf{Q}$ is  a bounded environment, and, $h_{\lambda} (\cdot)$ and $\phi(\cdot)$ 
 are  bounded, $\mathbf{W}_i(t,\mathbf{P}(t))$ is also bounded. 
 In addition, using the chain rule, we have
	\begin{align}\label{eq:38}
	\frac{\partial \mathbf{W}_{i}(t,\mathbf{P}(t))}{\partial \mathbf{p}_{j}(t)}=&\int_{\mathbf{Q}}\Big[\frac{\partial h_{\lambda}\!(t, R_i)}{\partial R_i}\ \! \frac{\partial R_i(\mathbf{q},\mathbf{P}(t))}{\partial \mathbf{p}_{j}(t)}\ \! 
	(\mathbf{q}-\mathbf{p}_{i}(t))^\mathrm{T}
	\notag
	\\
	&\qquad-h_{\lambda}\!(t, R_i)\frac{\partial\mathbf{p}_i}{\partial\mathbf{p}_j} \Big]\phi(\mathbf{q})	\mathrm{d}\mathbf{q}.
\end{align}	
$\frac{\partial h_{\lambda}\!(t, R_i)}{\partial R_i}$ is bounded according to {the definition of $h_{\lambda}$}. Moreover, $\frac{\partial R_i(\mathbf{q},\mathbf{P}(t))}{\partial \mathbf{p}_{j}(t)}$ is obtained as follows:
\begin{align}\label{r'}
	\frac{\partial R_i\big(\mathbf{q},\mathbf{P}(t)\big)}{\partial \mathbf{p}_{j}(t)}
	\!=\!& {\sum_{l=1}^{n}\frac{\partial \big(\lVert \mathbf{q}-\mathbf{p}_{i}(t) \rVert^2-\lVert \mathbf{q}-\mathbf{p}_{l}(t) \rVert^2\big)}{\partial \mathbf{p}_{j}(t)} }
	\notag\\ 
	&\quad\delta\big(\lVert \mathbf{q}-\mathbf{p}_{i}(t) \rVert^2-\lVert \mathbf{q}-\mathbf{p}_{l}(t) \rVert^2\big).
\end{align} 
Now, if we define 
$\mathcal{S}_{il}\delequal\{\mathbf{q}\in \mathbf{Q}|~\lVert \mathbf{q}-\mathbf{p}_{i}(t) \rVert=\lVert \mathbf{q}-\mathbf{p}_{l}(t)\rVert\}$,
which is the perpendicular bisector of  $\mathbf{p}_{i}$ and $\mathbf{p}_{l}$ in $\mathbf{Q}$, using Coarea formula \cite{zhang2021dirac} which states that
\begin{align}
	\int_{\mathbf{Q}} f(\mathbf{q})\delta(g(\mathbf{q}))\mathrm{d}\mathbf{q} = \int_{\{\mathbf{q}\in \mathbf{Q} | g(\mathbf{q})=0\}} \frac{f(\mathbf{s})}{\|\nabla g(\mathbf{s})\|}\mathrm{d}\mathbf{s},
\end{align}
we have for $j = i$:
\begin{align}
	\frac{\partial \mathbf{W}_i}{\partial \mathbf{p}_i} =& \sum_{k 
		\neq i}\int_{\mathcal{S}_{ik}} \frac{\partial h_{\lambda}\!(t, R_i)}{\partial R_i}\frac{(\mathbf{s}-\mathbf{p}_i)(\mathbf{s}-\mathbf{p}_i)^\mathrm{T}\phi(\mathbf{s})}{\|\mathbf{p}_k-\mathbf{p}_i\|}\mathrm{d}\mathbf{s} \notag
	\\
	&- \int_{\mathbf{Q}} h_{\lambda}(t, R_i)\phi\mathrm{d}\mathbf{q}\mathbf{I}_N,
	\label{s11}
\end{align}
and for $j \neq i$:
\begin{align}
	\frac{\partial \mathbf{W}_i}{\partial \mathbf{p}_j} = -\int_{\mathcal{S}_{ij}} \frac{\partial h_{\lambda}\!(t, R_i)}{\partial R_i}\frac{(\mathbf{s}-\mathbf{p}_j)(\mathbf{s}-\mathbf{p}_i)^\mathrm{T}\phi(\mathbf{s})}{\|\mathbf{p}_j-\mathbf{p}_i\|}\mathrm{d}\mathbf{s}.
	\label{s2}
\end{align}
Since $h_{\lambda}\!(t, R_i) \in (0,1]$, {then the second term of \eqref{s11} is bounded as follows}
\begin{equation}
	0<\int_{\mathbf{Q}}h_{\lambda}\!(t, R_i)\phi(\mathbf{q})	\mathrm{d}\mathbf{q}\leq\int_{\mathbf{Q}}\phi(\mathbf{q})	\mathrm{d}\mathbf{q}<\eta,
\end{equation}
and this completes the proof.
 \hfill $\blacksquare$
\endgroup

	The following theorem shows the  convergence of agents network to their optimal configuration using the proposed control law given by \eqref{f16}.
	\begin{theorem} 
		\textbf{(Known environment)}
		\label{theory:btf_known}
		\textit{	Assume that the density function of the environment is known and the dynamic of agents are defined by \eqref{f15_2}. If the control input of each agent is chosen as  \eqref{f16}, then  the control input of each agent converge to zero, and consequently the agents converge to a locally optimal coverage configuration.}
	\end{theorem}

	\textcolor{black}{\textbf{Proof.}}
	Consider the following Lyapunov-like function:
\begin{equation}
	V\big(t,\mathbf{P}(t)\big)=\hat{H}\big(t,\mathbf{P}(t)\big).
\end{equation}
By taking the time derivative of $V$, it follows that:
\begin{align}
	\dot{V}\big(t,\mathbf{P}(t)\big)&=\frac{\partial \hat{H}}{\partial t}+\sum_{i=1}^{n}\frac{\partial \hat{H}^\mathrm{T}\big(t,\mathbf{P}(t)\big)}{\partial \mathbf{p}_{i}(t)} \ \! \dot{\mathbf{p}}_{i}(t).
	\label{f18}
\end{align}
Using Lemma \ref{lemma:3} and \eqref{f16}, we can write:
\textcolor{black}{
\begin{align}
	\dot{V}\big(t,\mathbf{P}(t)\big)=&\frac{\partial \hat{H}}{\partial t}-\epsilon\sum_{i=1}^{n}\mathbf{W}_{i}^\mathrm{T}(t,\mathbf{P}(t))\mathbf{W}_{i}(t,\mathbf{P}(t)).
	\label{42}
\end{align}}
The derivative $\frac{\partial \hat{H}}{\partial t}$ is computed as follows:
\begin{align}
	\notag
	\frac{\partial \hat{H}}{\partial t}&=\!-\frac{1}{2}\!\sum_{i=1}^{n}\!\int _{Q}\!	\frac{\partial}{\partial t}\Big(\frac{R_i}{\lambda(t)}\Big)h_{\lambda}(t, R_i)\| \mathbf{q}\!-\!\mathbf{p}_{i}(t)\| ^{2}\!\phi \! \left( \mathbf{q}\right) \!\mathrm{d}\mathbf{q}
	\\
	&
	=\!\frac{1}{2}\!\sum_{i=1}^{n}\!\int _{Q}\!\frac{R_i\dot{\lambda}(t)}{\lambda^2(t)}h_{\lambda}(t, R_i)\| \mathbf{q}\!-\!\mathbf{p}_{i}(t)\| ^{2}\phi \! \left( \mathbf{q}\right) \!\mathrm{d}\mathbf{q}.				
\end{align}	
Hence, \eqref{42} can be written as follows:
\textcolor{black}{
\begin{align}
	\notag
	\!\dot{V}\big(t,\mathbf{P}(t)\big)\!\!=& \frac{1}{2}\!\sum_{i=1}^{n}\!\int _{Q}\!\frac{R_i\dot{\lambda}(t)}{\lambda^2(t)}h_{\lambda}(t, R_i)	\| \mathbf{q}\!-\!\mathbf{p}_{i}(t)\| ^{2}\phi \! \left( \mathbf{q}\right)\! \mathrm{d}\mathbf{q}
	\\
	&
	-\frac{1}{2}\sum_{i=1}^{n}2\epsilon \mathbf{W}^\mathrm{T}_i(t,\mathbf{P}(t))\mathbf{W}_i(t,\mathbf{P}(t)).
\end{align}}
Moreover, $\ddot{V}\big(t,\mathbf{P}(t)\big)$ is obtained as follows:
\textcolor{black}{
\begin{align} \label{eq:37_2}
	\notag
	\ddot{V}\big(t&,  \mathbf{P}(t)\big)\!=
	\\
	\notag
	&-\sum_{i=1}^{n}\!\Big(\!\int_{Q}\!\frac{R_i\dot{\lambda}(t)}{\lambda^2(t)} h_{\lambda}(t, R_i) 	(\mathbf{q} \!-\! \mathbf{p}_i(t)) 	
	\phi(\mathbf{q})\, \mathrm{d}\mathbf{q}\Big)^\mathrm{T}\dot{\mathbf{p}}_i(t)
	\\
	\notag
	&\!+\!\frac{1}{2}\!\sum_{i=1}^{n}\!\int_{Q}\!\Big(\frac{R_i\dot{\lambda}(t)}{\lambda^2(t)}\Big)^2 h_{\lambda}(t, R_i)\| \mathbf{q}\!-\!\mathbf{p}_{i}(t)\| ^{2} 	\phi \! \left( \mathbf{q}\right)\! \mathrm{d}\mathbf{q}
	\\
	\notag
	&
	\!+\!\frac{1}{2}\!\sum_{i=1}^{n}\!\int_{Q}\frac{R_i\ddot{\lambda}(t)\lambda^2(t)-2R_i\dot{\lambda}^2(t)\lambda(t)}{\lambda^4(t)} h_{\lambda}(t, R_i)
	\\
	&
	\notag
	\quad\qquad \| \mathbf{q}\!-\!\mathbf{p}_{i}(t)\| ^{2}\phi \! \left( \mathbf{q}\right) \!\mathrm{d}\mathbf{q}
	\\
	&\!
	-\!2\epsilon\!\sum_{i=1}^{n} \! \mathbf{W}^\mathrm{T}_i(t,\mathbf{P}(t))\frac{\partial \mathbf{W}_i(t,\mathbf{P}(t))}{\partial t}
	\notag
	\\
	&-\!2\epsilon^2\!\sum_{i=1}^{n} \! \mathbf{W}^\mathrm{T}_i(t,\mathbf{P}(t))\sum_{j=1}^{N}\frac{\partial \mathbf{W}_i(t,\mathbf{P}(t))}{\partial \mathbf{p}_j(t)}\mathbf{W}_j(t,\mathbf{P}(t))
	.
\end{align}}
\begingroup
\color{black} 

The first four terms in \eqref{eq:37_2} are of the form $\frac{R_i}{\lambda^n(t)}h_{\lambda}(R_i)=\frac{R_i}{\lambda^n(t)}\exp(-R_i/\lambda(t))$, $n\in\{2,3,4\}$. 
Since \(R_i\) takes integer values in \(\{0,1,\dots,n-1\}\), if \(R_i=0\), these four terms vanish identically. For  \(R_i\ge1\), by definition $c(t)\triangleq 1/\lambda(t)$, we can deduce that $\lim_{t\to\infty}\frac{R_i}{\lambda^n(t)}\,h_{\lambda}(R_i)=\lim_{C\to\infty}C^n e^{-C}=0$. Therefore, using  Lemma \ref{lemma:4}, we can conclude that $\ddot{V}(\cdot)$ is uniformly bounded.

Since $V(\cdot) > 0$ and $\dot{V}(\cdot) < 0$, $V(\cdot)$ decreases monotonically to a finite positive limit. By using Barbalat's lemma, $\dot{V}(\cdot) \to 0$, and from \eqref{42} and \eqref{f16}, we have 	$\lim_{t \to \infty} \mathbf{u}_i(t) = \lim_{t \to \infty} \mathbf{W}_i(t,\mathbf{P}(t)) = 0$, implying asymptotic convergence to the optimal coverage.  \hfill $\blacksquare$\endgroup

	\subsection{Unknown Environment}\label{sec4_2}
	In this section, assuming the environment or $\phi(\mathbf{q})$ is initially unknown, we propose a distributed adaptive control input for achieving optimal coverage configuration. Each agent estimates the density function online based on environment measurements and information exchange with other agents, then updates its position to converge to the optimal coverage configuration. For this purpose, consider following assumptions:
	\begin{assumption}
		\textbf{(Mathematical model for density function)}
		\label{assump1}
		The density function $\phi(\mathbf{q})$ is represented by a basis function scheme as
		\begin{equation}
			\label{a1}
			\phi(\mathbf{q})=\mathcal{K}^{\mathrm{T}}(\mathbf{q})\mathbf{a},
		\end{equation}
		where $\mathbf{a} \in {\mathbb{R}}_{\ge 0}^m$ is unknown  and $\mathcal{K}:\mathbf{Q} \to {\mathbb{R}}_{\ge 0}^m$ is known for each agent. Furthermore, ${a} ^{i}\geq a_\mathrm{min}$, $i =1,\dots, m$, and $a_\mathrm{min} \in \mathbb{R}_{\geq0}$ where ${a} ^{i}$ denotes the $i$-th element of the vector  $\mathbf{a}$.
	\end{assumption} 
	\begin{remark}
		This assumption is not limiting since any function with some smoothness requirements over a bounded domain can be approximated arbitrarily well by choosing some proper set of basis function \cite{sanner1991gaussian}.  
	\end{remark}
	\begin{assumption}\label{assump2}
		\textbf{(Network connection)}
		The network connection between agents is a connected network.
	\end{assumption}
	Since the parameter $\mathbf{a}$ is unknown, each agent needs to estimate this parameter to construct its control signal. The estimated parameter $\mathbf{a}$ and  $\phi(\mathbf{q})$ by the $i$-th agent at time $t$ are denoted by $\hat{\mathbf{a}}_{i}(t)$ and $\hat{\phi}_{i}(\mathbf{q},t)$, respectively. Hence, using \eqref{a1}, we have
	\begin{equation}
		\hat{\phi}_{i}(\mathbf{q},t)=\mathcal{K}^\mathrm{T}(\mathbf{q})\hat{\mathbf{a}}_{i}(t).
	\end{equation}
	{In this work, it is assumed that each agent is equipped with sensors capable of measuring $\phi(\mathbf{q})$ at its current position. In other words, the density function at the position $\mathbf{p}_{i}(t)$ of the $i$-th agent is directly accessible and known to that agent. This information allows each agent to make informed decisions based on the local density of the environment at its own location.}
	The estimated density function error denoted by $\tilde{\phi}_{i}(\mathbf{q},t)$, therefore, is calculated as 
	\begin{equation}
		\tilde{\phi}_{i}(\mathbf{q},t)\triangleq	\hat{\phi}_{i}(\mathbf{q},t)-\phi(\mathbf{q})=\mathcal{K}^\mathrm{T}(q)\tilde{\mathbf{a}}_{i}(t),
		\label{e2}
	\end{equation}
	where $\tilde{\mathbf{a}}_{i}(t)\triangleq\hat{\mathbf{a}}_{i}(t)-\mathbf{a}$.
	Similar to the known environment, the same control structure is proposed, where $\phi(\mathbf{q})$ is replaced by its  estimated value, $\hat{\phi}_{i}(\mathbf{q},t)$, i.e.
	\begin{equation}
		\mathbf{u}_{i}(t)=\epsilon\int_{\mathbf{Q}}h_{\lambda}\!(t, R_i)\big(\mathbf{q}-\mathbf{p}_{i}(t)\big)\hat{\phi}_{i}(\mathbf{q},t)\mathrm{d}\mathbf{q}.
		\label{f16_2}
	\end{equation}
	As we can see from 	\eqref{f16_2}, the estimation of density function has a vital role in the coverage problem. 
	\textcolor{black}{As previously stated, the $i$-th agent can sense and measure  $\phi(\mathbf{p}_{i}(t))$ and an approximation $\hat{\phi}_{i}(\mathbf{q},t)$ can be learned using sensor measurements and information shared from other agents during the agent's movement over time.}
	For this purpose, we propose  the modified version of the   adaptation law introduced in \cite{sch4} to update  $\hat{\mathbf{a}}_{i}(t)$ and consequently $\hat{\phi}_{i}(\mathbf{q},t)$ as follows:
	\begin{equation}
		\dot{\hat{\mathbf{a}}}_{i}(t)=\Gamma\big(\dot{\hat{\mathbf{a}}}_{\mathrm{pre}_{i}}(t)-\mathbf{I}_{\mathrm{proj}_{i}}(t)\dot{\hat{\mathbf{a}}}_{\mathrm{pre}_{i}}(t)\big),
		\label{aa2}
	\end{equation}
	where $\Gamma \in \mathbb{R}^{m\times m}$ is a diagonal positive-definite adaptation gain matrix and $\mathbf{I}_{\mathrm{proj}_{i}}(t)$ is a diagonal matrix  defined as follows
	\begin{equation}
		{I}_{\mathrm{proj}_{i}}^{j}(t)=
		\begin{cases}
			0, & \text{for} \quad \hat{{a}}_{i}^{j}(t)>{a}_{\mathrm{min}} , \\
			0, & \text{for} \quad \hat{{a}}_{i}^{j}(t)={a}_{\mathrm{min}}, \quad \text{and} \quad \dot{\hat{{a}}}_{\mathrm{pre}_{i}}^{j}(t) \ge 0,\\
			1, & \text{otherwise},
		\end{cases}
		\label{I_p}
	\end{equation}
	where ${I}_{\mathrm{proj}_{i}}^{j}(t)$  is  the $j$-th diagonal element for the matrix $\mathbf{I}_{\mathrm{proj}_{i}}(t)$.
	Furthermore, $\hat{{a}}_{i}^{j}(t)$ and $\dot{\hat{{a}}}_{\mathrm{pre}_{i}}^{j}(t)$ are the $j$-th element for $\hat{\mathbf{a}}(t)$ and $\dot{\hat{{\mathbf{a}}}}_{\mathrm{pre}_{i}}(t)$, respectively, and 
	\begin{align}\nonumber
		\dot{\hat{\mathbf{a}}}_{\mathrm{pre}_{i}}\!(t)=&-\mathbf{F}_{i}(t)\hat{\mathbf{a}}_{i}(t)-\gamma \big( \Lambda_i(t)\hat{\mathbf{a}}_{i}(t)-\Upsilon_i(t)\big)\\
		&-\zeta\sum_{j=1}^{n}l_{i,j}\big(\hat{\mathbf{a}}_{i}(t)-\hat{\mathbf{a}}_{j}(t)\big),
		\label{aa1}
	\end{align}
	where $\gamma>0$ is the scalar adaptation gain,  $\zeta>0$ is a consensus scalar gain, $\mathbf{F}_{i}(t)$ is a positive semi-definite matrix  defined as:
	\begin{align}
		\mathbf{F}_{i}(t)\triangleq&\epsilon\int_{\mathbf{Q}}\mathcal{K}(\mathbf{q})\big(\mathbf{q}-\mathbf{p}_{i}(t)\big)^\mathrm{T}h_{\lambda}\!(t, R_i)\mathrm{d}\mathbf{q} 
		\notag
		\\ 
		& \int_{\mathbf{Q}} \big(\mathbf{q}-\mathbf{p}_{i}(t)\big) \mathcal{K}^\mathrm{T}(\mathbf{q})h_{\lambda}\!(t, R_i)\mathrm{d}\mathbf{q},
	\end{align}
	and $\Upsilon_{i}(t)$ and $\Lambda_{i}(t)$ are defined as:
	\begin{align}\label{aa3}
		\Upsilon_{i}(t)&\triangleq\int_{0}^{t}w(\tau)\mathcal{K}\big(\mathbf{p}_{i}(\tau)\big)\phi\big(\mathbf{p}_{i}(\tau)\big)\mathrm{d}\tau,
		\\\label{eq:48}
		\Lambda_{i}(t)&\triangleq\int_{0}^{t}w(\tau)\mathcal{K}\big(\mathbf{p}_{i}(\tau)\big)\mathcal{K}^\mathrm{T}\big(\mathbf{p}_{i}(\tau)\big)\mathrm{d}\tau,
	\end{align}
	where $w(\tau)$ is a non-negative, selection-limited scalar weight function for data collection that keeps bounds on $\Lambda_{i}(t)$ and $\Upsilon_i(t)$.
	{\begin{assumption}\label{assump3}
			\textbf{(Sufficient path richness)}
			In this work it is assumed that the robots' paths are such that the matrix $\lim_{t\to \infty}\Lambda_{i}(t)$ is positive definite for all $i$.
		\end{assumption}
		Such an assumption is common in the coverage of unknown environment literature \cite{sch4}.}
	The weight $w(\tau)$ is chosen as a positive value for a certain period of time, after which its value becomes zero, i.e.
	\begin{equation}
		w(t)=
		\begin{cases}
			r, & \text{if} \quad t<\tau_w , \\
			0, & \text{otherwise},
		\end{cases}
	\end{equation}
	where  $r$ is a positive value applied during time $[0,\tau_w]$.
	The following theorem  proves the convergence of the algorithm for unknown environments. 
	%
	\begin{theorem}
		\textbf{(Unknown environment)}
		\label{theorem:btf_unknown}
		{Under Assumptions \ref{assump1} and \ref{assump2}, and the dynamic model defined by \eqref{f15_2}, the control input of each agent tends to zero, i.e., the agents converge to a optimal coverage configuration. Moreover, if Assumption \ref{assump3} holds, the estimated density function error tends to its global optimum as $t\to \infty$.}
	\end{theorem}

	Consider the following 	Lyapunov-like function:
	\begin{equation}
		V_2\big(t,\mathbf{P}(t),\tilde{\mathbf{A}}(t)\big)=\hat{H}\big(t,\mathbf{P}(t)\big)+\frac{1}{2}\sum_{i=1}^{n}\tilde{\mathbf{a}}_{i}^\mathrm{T}(t)\Gamma^{-1}\tilde{\mathbf{a}}_{i}(t).
	\end{equation}
	where $\tilde{\mathbf{A}}(t)\triangleq\{\tilde{\mathbf{a}}_{1}(t),\dots,\tilde{\mathbf{a}}_{n}(t)\}$.
		The time derivative of $V_2(\cdot)$ is as follows:
	\begin{equation}
		\dot{V}_2(\cdot)=\frac{\partial \hat{H}}{\partial t}+\sum_{i=1}^{n} \Big[\frac{\partial \hat{H}^\mathrm{T}(t)}{\partial \mathbf{p}_{i}(t)} \ \! \dot{\mathbf{p}}_{i}(t)+\tilde{\mathbf{a}}_{i}^\mathrm{T}(t)\Gamma^{-1}\dot{\tilde{\mathbf{a}}}_{i}(t)\Big].
	\end{equation}
	Using Lemma \ref{lemma:3}, one can write:
	\begin{align}
	\dot{V}_2(\cdot)\!=
	&
	\notag
	\!\sum_{i=1}^{n} \!\Big[\frac{1}{2}\int _{Q}\frac{R_i\dot{\lambda}(t)}{\lambda^2(t)}h_{\lambda}\!(t, R_i)	\| \mathbf{q}-\mathbf{p}_{i}(t)\| ^{2}\phi \left( \mathbf{q}\right) d\mathbf{q}	
	\\
	&
	-\!\int_{\mathbf{Q}}\!\!\! h_{\lambda}\!(t, R_i)(\mathbf{q}-\mathbf{p}_{i}(t))\!^\mathrm{T}(\hat{\phi}_i(\mathbf{q},t)-\tilde{\phi}_i(\mathbf{q},t))\dot{\mathbf{p}}_i(t)\mathrm{d}\mathbf{q}
	\notag
	\\
	&
	+\tilde{\mathbf{a}}_{i}^\mathrm{T}(t)\Gamma^{-1}\dot{\hat{\mathbf{a}}}_{i}(t)\Big].
\end{align}
	Finally, using \eqref{aa3}, and by substituting $\phi(\mathbf{q})$, $\dot{\mathbf{p}}_{i}(t)$ and $\dot{\hat{\mathbf{a}}}_{i}(t)$ from  \eqref{e2},  \eqref{f16} and \eqref{aa2}, respectively, in the above equation one can write:
	\begin{align}
	\dot{V}_2(\cdot)\!&=
	\notag
	\!\frac{1}{2}\sum_{i=1}^{n} \!\int _{Q}\frac{R_i\dot{\lambda}(t)}{\lambda^2(t)}h_{\lambda}(t, R_i)	\| \mathbf{q}-\mathbf{p}_{i}(t)\| ^{2}\phi \! \left( \mathbf{q}\right)\! \mathrm{d}\mathbf{q}	
	\\
	&  +\sum_{i=1}^{n} \Big[-\lVert \int_{\mathbf{Q}} h_{\lambda}(t, R_i)(\mathbf{q}-\mathbf{p}_{i}(t))	\hat{\phi}_i(\mathbf{q},t) \mathrm{d}\mathbf{q}\rVert^2
	\notag
	\\
	&
	-\tilde{\mathbf{a}}_{i}^\mathrm{T}(t)\gamma\Lambda_{i}(t)\tilde{\mathbf{a}}_{i}(t)	-\tilde{\mathbf{a}}^\mathrm{T}_{i}(t)	\mathrm{\mathbf{I}}_{\mathrm{proj}_{i}}\!(t)\dot{\hat{\mathbf{a}}}_{\mathrm{pre}_{i}}\!(t) \Big]
	\notag
	\\
	&
	-\!\sum_{i=1}^{n}\!\tilde{\mathbf{a}}_{i}^\mathrm{T}(t)\zeta\!\sum_{j=1}^{n}\! l_{i,j}\big[\hat{\mathbf{a}}_{i}(t)\!-\!\hat{\mathbf{a}}_{j}(t)\big].
	\label{pr1}
\end{align} 
	For ease of referencing, 
	denote the five terms in the right-hand side of \eqref{pr1} as $\varrho_1(t)$, $\varrho_2(t)$, $\varrho_3(t)$, $\varrho_4(t)$, and $\varrho_5(t)$, respectively. It follows that
	$\varrho_5(t)=-\zeta\sum_{j=1}^{m}\tilde{\mathbf{\Omega}}_j^\mathrm{T}(t)\mathbf{L}\hat{\mathbf{\Omega}}_j(t)$, where $\mathbf{\Omega}_j(t)\triangleq{a}^j\mathbf{1}_{n \times 1}$,  $\hat{\mathbf{\Omega}}_j(t) \triangleq [\hat{{a}}_1^j(t),\dots,\hat{a}_n^j(t)]^\mathrm{T}$ and $\tilde{\mathbf{\Omega}}_j(t) \triangleq\hat{\mathbf{\Omega}}_j(t)-\mathbf{\Omega}_j(t)$. 
	Based on the properties of the Laplacian matrix $\mathbf{L}$, $\mathbf{\Omega}_j^\mathrm{T}\mathbf{L}=a^j\mathbf{1}^\mathrm{T}_{n \times 1}\mathbf{L}=0 ,\forall j$. Therefore,
	\begin{equation}
		-\zeta\sum_{j=1}^{m}\tilde{\mathbf{\Omega}}_j^\mathrm{T}(t)\mathbf{L}\hat{\mathbf{\Omega}}_j(t)=-\zeta\sum_{j=1}^{m}\hat{\mathbf{\Omega}}_j^\mathrm{T}(t)\mathbf{L}\hat{\mathbf{\Omega}}_j(t).
		\label{lap_1}
	\end{equation}	
	It is clear that $\varrho_1(t)< 0$, $\varrho_2(t)\le 0$, and
according to the definition of $w(t)$ and $\mathbf{L}$, $\varrho_3(t)$ and $\varrho_5(t)$ are also non-positive.
Using \eqref{I_p}, if $\hat{{a}}_{i}^{j}(t)>{a}_{\mathrm{min}}$ or $\hat{{a}}_{i}^{j}(t)={a}_{\mathrm{min}}$ and $\dot{\hat{{a}}}_{\mathrm{pre}_{i}}^{j}(t) \ge 0$, then $\mathrm{{I}}_{\mathrm{proj}_{i}}^{j}(t)=0$,  and the corresponding term is vanished.
On the other hand, if $\hat{{a}}_{i}^{j}(t)={a}_{\mathrm{min}}$ and $\dot{\hat{{a}}}_{\mathrm{pre}_{i}}^{j}(t) < 0$, then $\mathrm{{I}}_{\mathrm{proj}_{i}}^{j}(t)=1$ and we observe that $\tilde{{a}}_{i}^{j}(t)=\hat{{a}}_{i}^{j}(t)-{a}^{j} \le 0$, since ${a}^{j}\geq {a}_\mathrm{min}$. 
Therefore,  $\varrho_4(t)\le 0$, and, consequently, $\dot{V}_2(\cdot)< 0$.
It can be shown that $\ddot{V}_2(\cdot)$ is bounded and $\dot{V}_2(\cdot)$ is therefore uniformly continuous.
Since  $V_2(\cdot)>0$ (is lower bounded) and $\dot{V}_2(\cdot)< 0$, then $V_2(\cdot)$ has a finite limit. Therefore, using  Barbalat's lemma, $\dot{V}_2(\cdot)\to 0$, and as a result  $\varrho_i(t)\to 0$, $i=1, \dots,5$. Hence, it follows that:
\begin{align}
	\nonumber\lim_{t\to \infty} \int_{\mathbf{Q}}h_{\lambda}\big(t,R_i(\mathbf{q},\mathbf{P}(t))\big)\big(\mathbf{q}-\mathbf{p}_{i}(t)\big)\hat{\phi}_{i}(\mathbf{q},t)\mathrm{d}\mathbf{q}=0.
	\label{p_un1}
\end{align}	
Additionally, based on Assumption 3, since $\dot{V_2}\to 0$ and $\lim_{t\to \infty}\Lambda_{i}(t)$ is positive definite, then $\tilde{\mathbf{a}}_{i}\to 0$. Utilizing (\ref{e2}), it follows that $\lim_{t\to \infty} \tilde{\phi}_{i}(\mathbf{q},t) =0$. Thus, the estimation of the agent $i$ converges to the  true approximation of the density function, $\phi(\mathbf{q})$. \hfill $\blacksquare$
	
	{\begin{remark}
			If Assumption \ref{assump3} does not hold, the agents' estimation of the environment does not converge to its global optimum, meaning that the agents do not fully identify the environment. Although the control law of agents still converges to a local optimum, this local optimum might not correspond to the optimum of the real environment, but rather to the one that each agent has identified, which might be incorrect.
	\end{remark}}
	
	\section{Simulation Results}\label{s_5}
	In this section, the simulation results of the proposed coverage method in known and unknown environments are presented. In order to evaluate the effectiveness of the proposed method, various scenarios are conducted. 
	Furthermore, to show the performance of the proposed method in comparison with  other approaches, the proposed method is compared with presented methods in \cite{bullo4} and \cite{sch4} in known and unknown environments, respectively, based on the cost function given by \eqref{f1}.
	It should be noted that, the initial position of  agents in  comparative studies is considered to be the same. {We included extra scenarios not detailed in the paper due to brevity. However, one can access videos of all scenarios through the following link {\textit{\urlstyle{rm}\url{https://www.youtube.com/watch?v=6QcD1WQSBZY}}}.}
	The following assumptions are considered in this section:	
	\begin{itemize}
		
		\item The environment $\mathbf{Q}$ is taken to be  a unit square and it  is  divided into an even $5\times5$ grid. Each truncated Gaussian mean $\mu_j$ is set so that each of  $25$ Gaussian functions is centered at its corresponding grid square.
		\item To ensure a fair comparison between the performance of our algorithm and the Voronoi method, the control gains $\epsilon$ and $k_{\mathrm{prop}}$ are regulated so that both algorithms operate within the same control input range.
		\item \textcolor{black}{For implementation purposes, the algorithm is discretized with a sampling time of $0.01\rm s$.}
		
		\item The density function of the environment is parameterized based on Assumption \ref{assump1} and with the help of $25$ truncated Gaussian functions (i.e. $\mathcal{K}(\mathbf{q})=[ \mathcal{K}_1(\mathbf{q}),\dots,\mathcal{K}_{25}(\mathbf{q})]^\mathrm{T}$) as follows:
		\begin{equation}
			\label{a4}
			{\mathcal{K}}_{j}(\mathbf{q})\triangleq
			\begin{cases}
				G_j(\mathbf{q})-G_{\mathrm{trunc}}, & \text{if} \quad \lVert \mathbf{q}-\mu_j \rVert<\rho_{\mathrm{trunc}}, \\
				0 ,             & \text{otherwise},
			\end{cases}
		\end{equation}
		where
		\begin{align}
			\label{a3}
			G_j(\mathbf{q})&\triangleq\frac{1}{\sigma \sqrt{2 \pi}} \text{exp}\Big\{-\frac{\lVert \mathbf{q}-\mathbf{\mu}_j \rVert^2}{2 \sigma^2}\Big\},\\	
			\label{a2}
			G_{\mathrm{trunc}}&\triangleq\frac{1}{\sigma \sqrt{2 \pi}} \text{exp}\Big\{-\frac{\rho_{\mathrm{trunc}}^{2}}{2 \sigma^2}\Big\},
		\end{align}
		and $\mu_j$ is the mean of $G_j$ and $\sigma$ is the standard deviation of Gaussian functions. The parameter $\rho_{\mathrm{trunc}}$ is the radius of a circle to the center of $\mu_j$. 
	\end{itemize}
	\begin{table}[h]
		\centering
		\caption{Parameters of the proposed method for simulation scenarios of the known environment.}
		\label{t2}
		\begin{tabular}{ccccccccccc}
			\hline
			Parameter	&$\epsilon$  &$\lambda_\mathrm{f}$ &$\lambda_{\mathrm{s}}$&$\alpha$&$\rho_{\mathrm{trunc}}$\\ \hline
			Value &$0.1$ &$2\times10^{-3}$ 	& $4$	& $25\times10^{+3}$	& $0.2$
		\end{tabular}
	\end{table}
	
	\subsection{Known environments}\label{sec4}
	\textcolor{black}{
		In this subsection, the performance of the proposed method is evaluated and compared to the Voronoi-based method in \cite{bullo4} in a known environment. The detailed information of the selected parameters for these scenarios are given in TABLE \ref{t2}. 
	}

\begin{figure}[pt]
		\centering
		\hspace*{-.4cm}
		\includegraphics[scale=0.36,trim={2.5cm .5cm 2cm .5cm},clip]{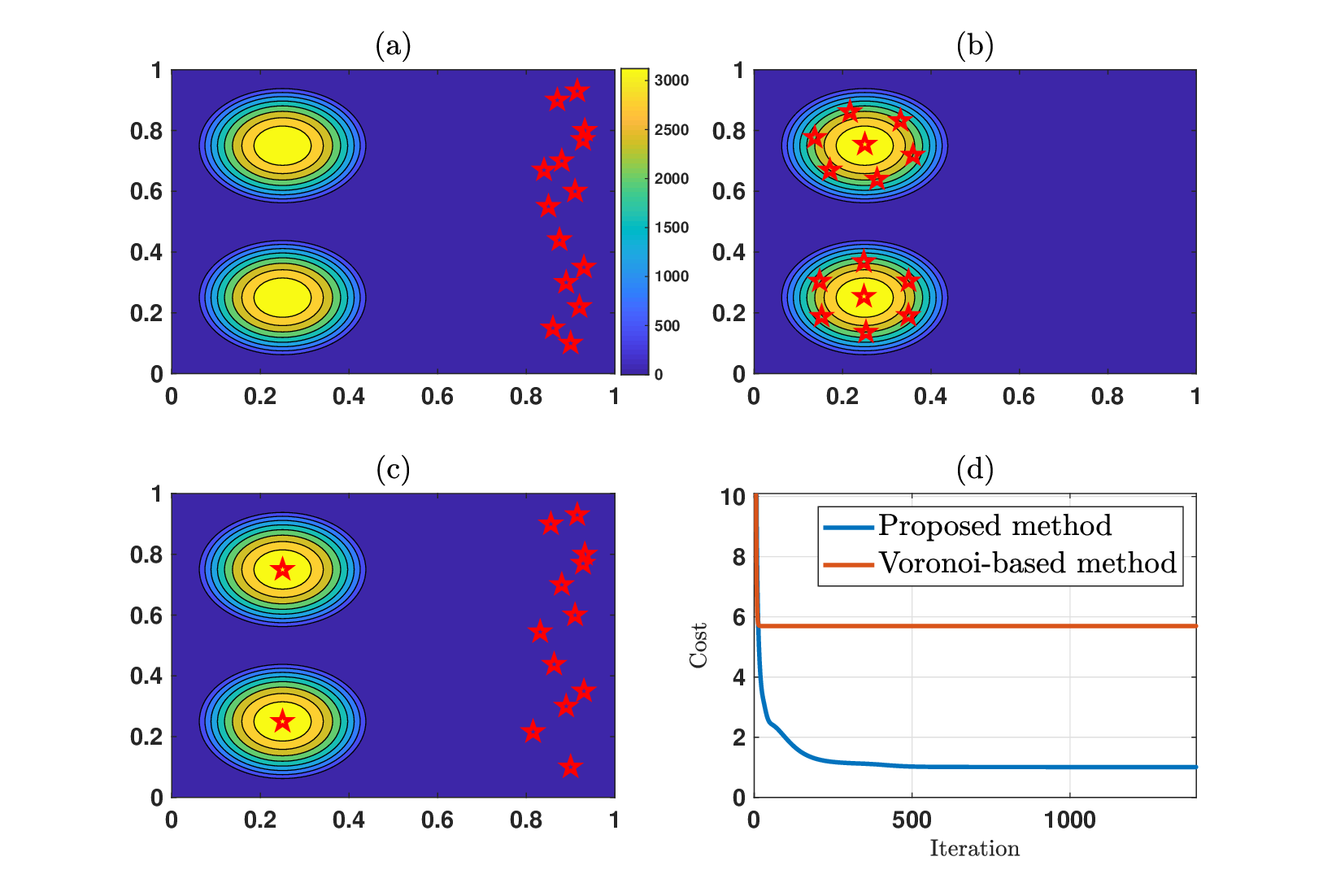}
		\caption{Simulation results of the third scenario for known environment containing separated information areas: (a) Initial configuration of the agents. (b) Final configuration using the proposed method. (c) Final configuration using the Voronoi-based method \cite{bullo4}. (d) The value of the cost function at each step.} 
		\label{fig7}
	\end{figure}

	In \textcolor{black}{this} scenario, the performance of the proposed method is investigated in the environment with separated information areas.
	For this purpose, two separated information areas are modeled using the following vector $\mathbf{a}$:
	\begin{equation}	
		\mathbf{a}_{25 \times 1}\triangleq
		\begin{cases}
			{a}^{j}=29960, & \text{for} \quad j=7 \quad \text{and} \quad j=9 , \\
			{a}^{j}=0,     & \text{otherwise}.
		\end{cases}
		\label{separate}
	\end{equation}
	\textcolor{black}{The proposed method, as shown in Fig. \ref{fig7}(c), distributes agents across two regions with higher density functions, while the Voronoi-based method introduced by \cite{bullo4} places most of the  agents in a area with zero density function.} Clearly, as shown in Fig. \ref{fig7}(d),  the cost generated by the proposed method is much less than the Voronoi-based method. The results of the considered scenarios show that the proposed method effectively distributes agents in the environment,  incurs a final cost that is $80\%$ lower than the Voronoi-based approach.
	
	\subsection{Unknown environments}	
\begin{figure}[pt]
		\centering
		\hspace*{-.4cm}
		\includegraphics[scale=0.36,trim={2.5cm .5cm 2cm .5cm},clip]{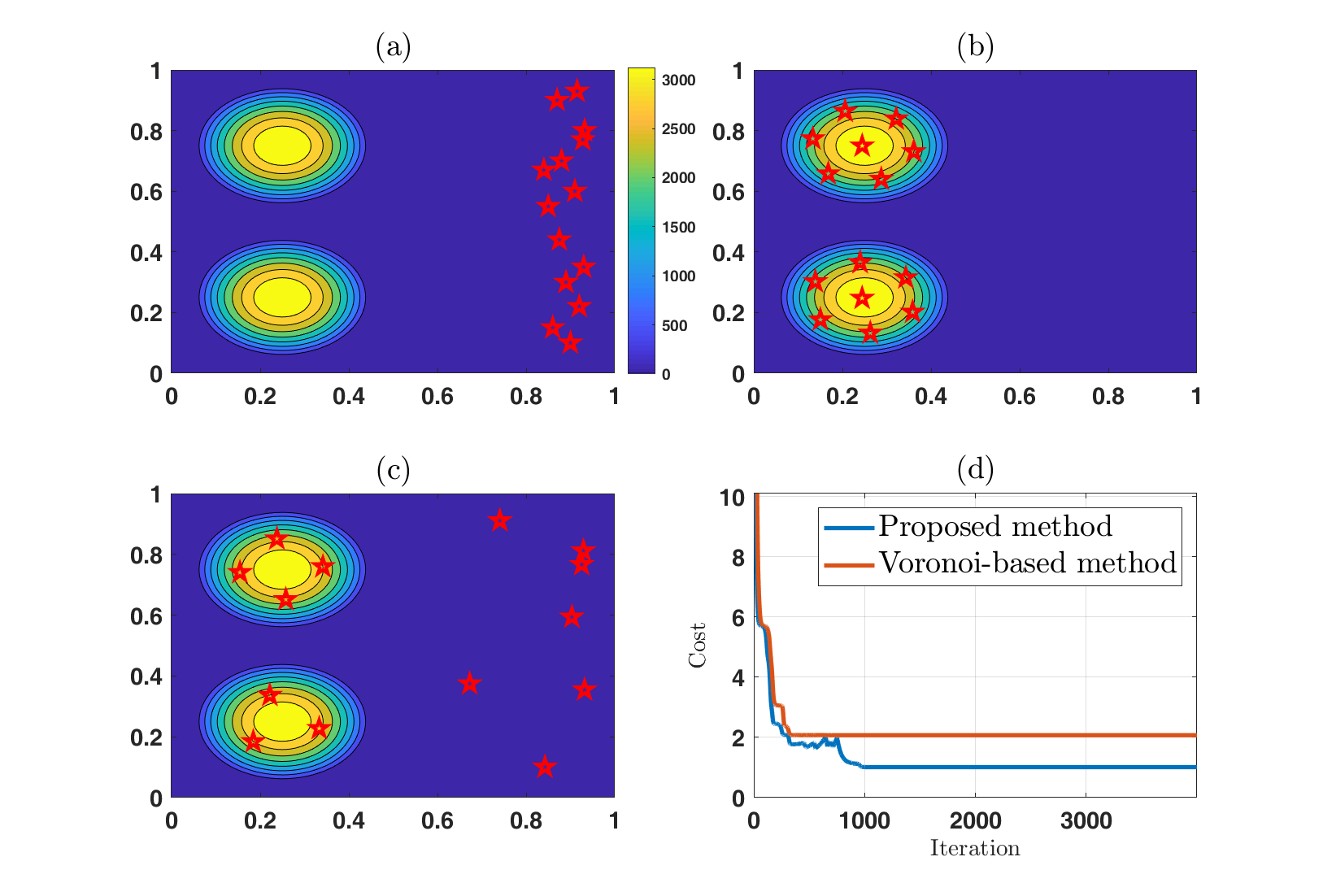}
		\caption{Simulation results of scenario for unknown environment containing separated information areas: (a) Initial configuration of the agents. (b) Final configuration using the proposed method. (c) Final configuration using the Voronoi-based method \cite{sch4} . (d) The value of the cost function at each step.} 
		\label{u1}
	\end{figure}

	In this subsection,  the performance of the proposed method in comparison with the Voronoi-based method introduced by \cite{sch4} in an unknown environment with separated information areas  is investigated. For this purpose, two separated information areas  modeled by \eqref{separate} are considered. The detailed information of the  parameters are given in TABLE \ref{t1}. The initial  configuration of agents is  shown in Fig. \ref{u1}(a).
	
	\begin{table}[t]
		\centering
		\caption{ Parameters of the proposed method for simulation scenario of the unknown environment.
		}
		\label{t1}
		\scalebox{.8}{
			\begin{tabular}{cccccccccccc}
				\hline
				Parameter	&$\gamma$  &$\zeta$ &$l_{i,j}$ &$\epsilon$ &$w$ &$\mathbf{\hat{a}}_{i_{\mathrm{initial}}}$ &$\mathbf{a}_{\min}$	&$\rho_{\mathrm{trunc}}$\\ \hline
				Value &$0.14$ &$0.01$ &$1$ &$0.9$ &$180$ &$400 \times \mathbf{1}_{25 \times 1}$ &$\mathbf{0}_{25 \times 1}$ & $0.2$
		\end{tabular}}
	\end{table}

	\textcolor{black}{
		Note that, to facilitate learning the distribution of information in the environment, initially, the Gaussian peaks centered in all cells are assumed to be the same. This causes the agents to have a tendency to cover all regions and spread evenly throughout the environment, resulting in better learning. This is the reason why the Voronoi algorithm performs better in the unknown environment compared to the known scenario. In this case, the agents initially assume that there is information distributed throughout the environment, prompting them to move. These movements prevent the agents from getting stuck in zero information areas.}
	\textcolor{black}{Moreover, unlike the known scenarios where $\lambda(\cdot)$ started from a higher value and monotonically decreased, here we start with a relatively low value for $\lambda(\cdot)$ in the early stages and gradually increase it until the agents have learned the environment, as indicated by a small relative change in $\hat{\mathbf{a}}$. 
		The formula for the relative change in the objective function (RCOV) value is defined as follows:
		\begin{equation}\label{RCOV}
			RCOV = \frac{{|f({x_{k + 1}}) - f({x_k})|}}{{|f({x_k})|}} \times 100\%,
		\end{equation}
		where $f({x_k})$ is the value of the objective function at the $k$-th iteration.
		Once a threshold of $1\%$ is reached, we allow $\lambda(\cdot)$ to decrease monotonically, similar to the known scenario. This prevents the agents from concentrating in the center of information in the early stages and allows them to spread and learn the environment more effectively.}

\begin{figure}[t]
		\centering
		\subfigure[]{\includegraphics[height=3.2cm,width=4.2cm]{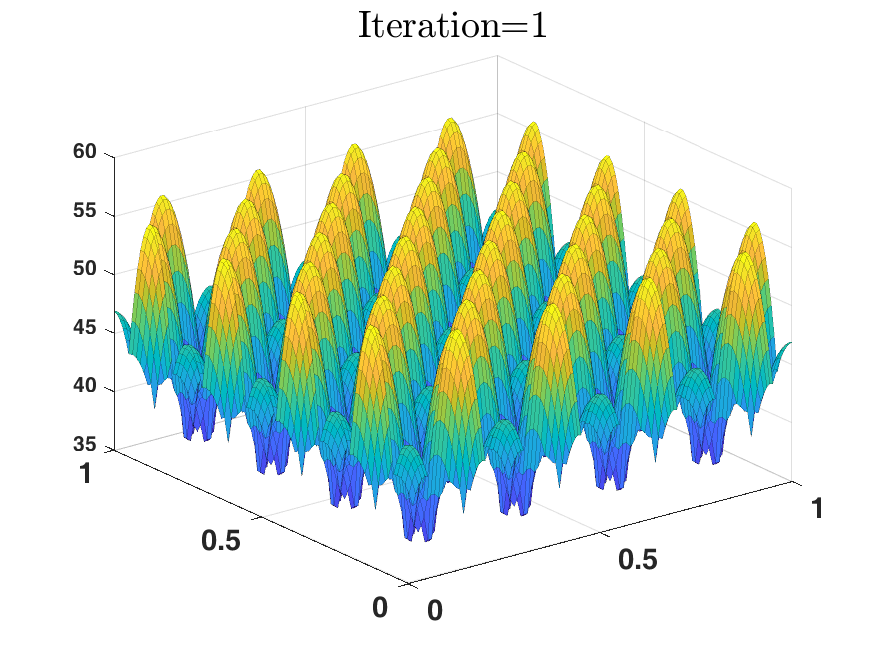}}\hfil
		\subfigure[]{\includegraphics[height=3.2cm,width=4.2cm]{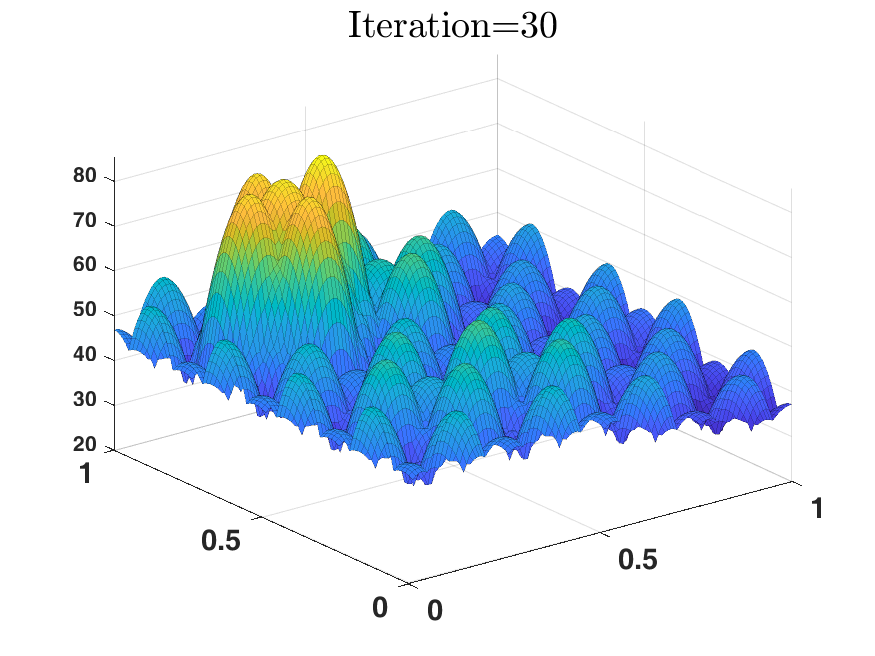}}
		\hfil
		\subfigure[]{\includegraphics[height=3.2cm,width=4.2cm]{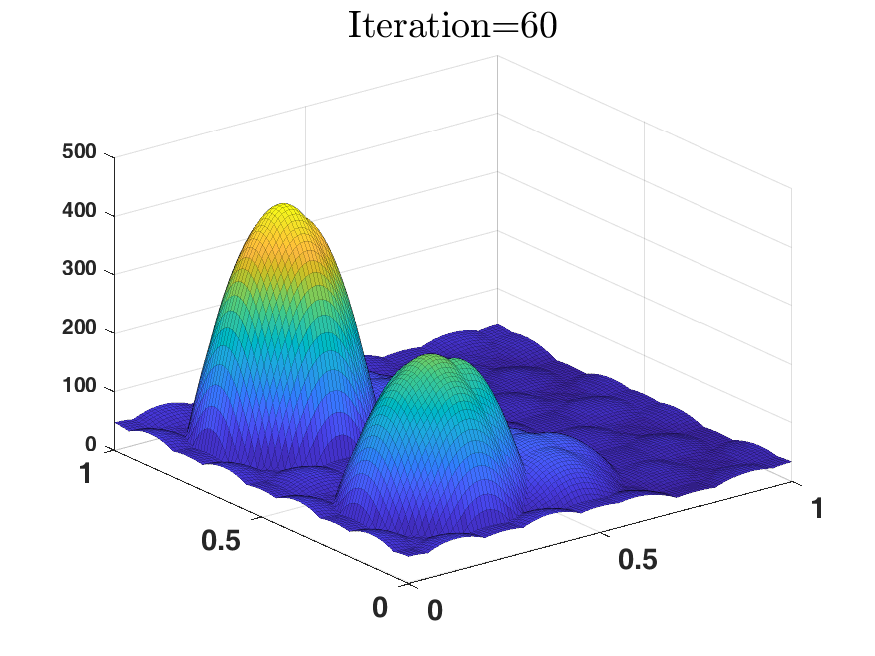}}\hfil
		\subfigure[]{\includegraphics[height=3.2cm,width=4.2cm]{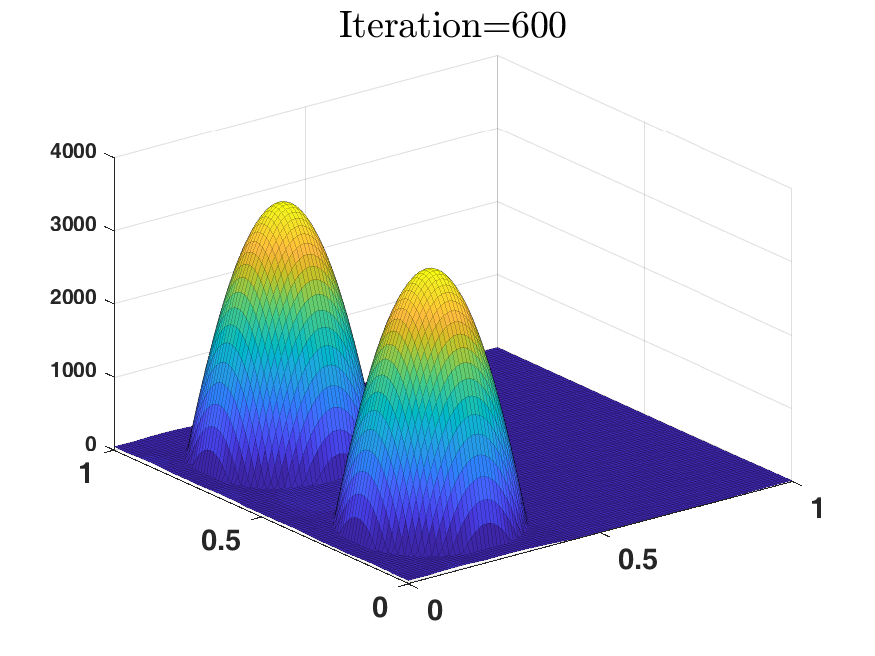}}
		\caption{Environment estimation in iterations 1, 30, 60 and 600 are shown in figures (a), (b), (c) and (d) respectively.}
		\label{vfig2}
	\end{figure}

	The simulated results of the proposed method and the Voronoi-based method are shown in Fig. \ref{u1}(b) and Fig. \ref{u1}(c), respectively. It is clear that the proposed method is effectively distributed the agents among the information areas compared to the Voronoi-based method.
	The final cost value of the proposed method, as shown in Fig. \ref{u1}(d), is $50\%$ lower than the Voronoi-based method. In addition, the estimated density function of  the  environment in different iterations is shown in Fig. \ref{vfig2}. {It's worth noting that the final cost and configuration of agents using our method are identical. This indicates that our method is less sensitive to local minima along the way.}
	\subsection{Discussion}
	This section provides a comprehensive discussion by exploring key aspects, including computational complexity, and convergence rate, to assess the algorithm's efficiency, effectiveness, and suitability for various real-world applications. Finally, to conclude our examination, we investigate the limitations encountered and discuss future research directions.
	
	\subsubsection{Complexity analysis}
	In this section, we delve into the computational complexity of our proposed algorithm, comparing it to that of the Voronoi coverage algorithm. {It is worth noting that, for implementation purposes, we consider the discrete form of (\ref{f1_45}) where  the environment $Q$ is discretized into $N$ points.}
	
	In the Voronoi coverage algorithm, the primary computation of the discrete control signal entails two main steps. First, there is a summation over all points in set $Q$, and then the computation of ${h}_i(\cdot)$ values. 
	Calculating ${h}_i(\cdot)$ involves computing the distance from the point $q$ to all $n$ agents, which amounts to order $n$ operation, and comparing these distances between the current agent and the point to all other computed distances to determine the value of $h_i(\cdot)$, which involves $n-1$ comparisons. This results in a computation of complexity order $N(2n-1)$. Given that $N \gg 2n-1$, the complexity of this algorithm can be expressed as $\mathcal{O}(N)$.
	
	Likewise the computational complexity of our proposed method depends on a summation over all points in set $Q$. However, we need to deal with $h_{\lambda}(\cdot)$ instead of $h_i(\cdot)$. Similar to $h_i(\cdot)$, computing the values for $h_{\lambda}(\cdot)$ requires us to initially determine the distance from point $q$ to all $n$ agents, which amounts to order $n$ operation. However, when constructing the ranking function (\ref{f3}), we employ a Quicksort algorithm as described in \cite{hoare1962quicksort,Quicksiam}, which in its worst-case scenario results in a complexity order of $n\log(n)$. Therefore, the complexity of our algorithm will be of order $N(n\log(n)+n)$. Given that $N \gg n(\log(n)+1)$, the computational complexity of our algorithm can also be expressed as $\mathcal{O}(N)$. Therefore, both algorithms exhibit the same order of computational complexity.
	
	\begin{table}[]
		\centering
		\caption{{The number of iterations required for convergence in scenario 3 for various $\epsilon$ values}}
			\scalebox{1}{\begin{tabular}{llllll}\cline{1-6}
			$\epsilon$& 0.10 &0.13 &0.15 &0.2 &0.3  \\\cline{1-6}
			Number of Iteration &235 &40  & 30 &28 &20
		\end{tabular}}\label{tableeps}
	\end{table}
	\subsubsection{Input range and convergence rate}
	The algorithm's convergence rate, crucial for assessing efficiency, is closely linked to the agents' operating input range. Our approach allows flexible adjustment of the control signal's amplitude range by modifying a positive scalar value, $\epsilon$, acting as a gain factor. Higher $\epsilon$ values expand the permissible control inputs, enabling quicker agent position adjustments in dynamic conditions.
	
	{
		To illustrate this fact, the RCOV given by (\ref{RCOV}) is used as a metric to measure the convergence of our algorithm. In this work, a threshold of $0.1\%$ is employed below which the algorithm is considered to have converged. TABLE \ref{tableeps} shows the the number of iterations required for our algorithm to converge for various values of $\epsilon$ in the third scenario.
		As can be seen from TABLE \ref{tableeps}, higher gain values effectively accelerate the convergence of the algorithm.}
	However, it is crucial to note that the choice of gain $\epsilon$ should be made carefully to ensure that agents remain within the convex environment. This constraint is necessary to facilitate the minimization of the cost function.

	\subsubsection{Limitations and future works}
	This work exclusively deals with a simplified representation of agents, considering them as point entities characterized by single integrator dynamics. We have not accounted for the agents' specific longitudinal and lateral  dimensions in this work. Additionally, we have assumed that these agents are holonomic, meaning they can move freely in any direction without any constraints.  Moreover, the environment is assumed to be obstacle-free.
	
	However, there are several avenues for future research in this field. Firstly, researchers can delve into more complex vehicle dynamics, considering real-world scenarios where agents may have specific sizes and limitations in their movements, such as non-holonomic constraints. This would lead to a more accurate modeling of the agents' behavior in practical situations.
	{Secondly, addressing the coverage problem in environments containing obstacles presents an intriguing challenge. In many real-world applications, agents must navigate around obstacles to effectively cover the entire area of interest. Investigating strategies and algorithms that allow agents to intelligently avoid obstacles while optimizing coverage could significantly advance this field of research.}

	\section{Conclusion}\label{s_6}
	In this work, we presented a novel approach to tackle the challenges associated with coverage control in multi-agent systems by employing a unique cost function that progressively aligns with the conventional Voronoi-based cost function over time, leading to enhanced performance. 
	Theoretical analyses demonstrated the asymptotic convergence of the algorithm. 
	Furthermore, our technique proves to be suitable for environments with both known and unknown information distributions. Finally, through rigorous simulations, we have showcased the efficacy of our proposed method, comparing it with Voronoi-based algorithms.

	\bibliographystyle{plain}        
	\bibliography{doc}           

\begin{thebibliography}{10}

\bibitem{roi3}
Farshid Abbasi, Afshin Mesbahi, and Javad~Mohammadpour Velni.
\newblock A team-based approach for coverage control of moving sensor networks.
\newblock {\em Automatica}, 81:342--349, 2017.

\bibitem{DR1}
Rihab Abdul~Razak, Sukumar Srikant, and Hoam Chung.
\newblock Decentralized and adaptive control of multiple nonholonomic robots
  for sensing coverage.
\newblock {\em International Journal of Robust and Nonlinear Control},
  28(6):2636--2650, 2018.

\bibitem{est1}
Alessia Benevento, Mar{\'\i}a Santos, Giuseppe Notarstefano, Kamran Paynabar,
  Matthieu Bloch, and Magnus Egerstedt.
\newblock Multi-robot coordination for estimation and coverage of unknown
  spatial fields.
\newblock In {\em {IEEE} {I}nternational {C}onference on {R}obotics and
  {A}utomation ({ICRA})}, pages 7740--7746, 2020.

\bibitem{G_caicedo2008coverage}
Carlos~Humberto Caicedo-N{\'u}nez and Milos Zefran.
\newblock A coverage algorithm for a class of non-convex regions.
\newblock In {\em 2008 47th IEEE Conference on Decision and Control}, pages
  4244--4249. IEEE, 2008.

\bibitem{BH3}
Andrea Carron and Melanie~N Zeilinger.
\newblock Model predictive coverage control.
\newblock {\em IFAC-PapersOnLine}, 53(2):6107--6112, 2020.

\bibitem{bullo4}
Jorge Cortes, Sonia Martinez, Timur Karatas, and Francesco Bullo.
\newblock Coverage control for mobile sensing networks.
\newblock {\em IEEE Transactions on robotics and Automation}, 20(2):243--255,
  2004.

\bibitem{hoare1962quicksort}
Charles~AR Hoare.
\newblock Quicksort.
\newblock {\em The computer journal}, 5(1):10--16, 1962.

\bibitem{C_hokayem2007persistent}
Peter~F Hokayem, Dusan Stipanovic, and Mark~W Spong.
\newblock On persistent coverage control.
\newblock In {\em 2007 46th IEEE Conference on Decision and Control}, pages
  6130--6135. IEEE, 2007.

\bibitem{kantaros2015distributed}
Yiannis Kantaros, Michalis Thanou, and Anthony Tzes.
\newblock Distributed coverage control for concave areas by a heterogeneous
  robot--swarm with visibility sensing constraints.
\newblock {\em Automatica}, 53:195--207, 2015.

\bibitem{D_leonard2013nonuniform}
Naomi~Ehrich Leonard and Alex Olshevsky.
\newblock Nonuniform coverage control on the line.
\newblock {\em IEEE Transactions on Automatic Control}, 58(11):2743--2755,
  2013.

\bibitem{B_li2005distributed}
Wei Li and Christos~G Cassandras.
\newblock Distributed cooperative coverage control of sensor networks.
\newblock In {\em Proceedings of the 44th IEEE Conference on Decision and
  Control}, pages 2542--2547. IEEE, 2005.

\bibitem{pso1}
Haifeng Ling, Tao Zhu, Weixiong He, Hongchuan Luo, Qing Wang, and Yi~Jiang.
\newblock Coverage optimization of sensors under multiple constraints using the
  improved {PSO} algorithm.
\newblock {\em Mathematical Problems in Engineering}, 2020.

\bibitem{Quicksiam}
Conrado Mart\'{\i}nez and Salvador Roura.
\newblock Optimal sampling strategies in quicksort and quickselect.
\newblock {\em SIAM Journal on Computing}, 31(3):683--705, 2001.

\bibitem{F_rickenbach2024active}
Rahel Rickenbach, Johannes K{\"o}hler, Anna Scampicchio, Melanie~N Zeilinger,
  and Andrea Carron.
\newblock Active learning-based model predictive coverage control.
\newblock {\em IEEE Transactions on Automatic Control}, 2024.

\bibitem{selft}
Erick~J Rodr{\'\i}guez-Seda, Xiaotian Xu, Josep~M Olm, Arnau D{\`o}ria-Cerezo,
  and Yancy Diaz-Mercado.
\newblock Self-triggered coverage control for mobile sensors.
\newblock {\em IEEE Transactions on Robotics}, 2022.

\bibitem{sanner1991gaussian}
Robert~M Sanner and Jean-Jacques~E Slotine.
\newblock Gaussian networks for direct adaptive control.
\newblock In {\em 1991 American control conference}, pages 2153--2159. IEEE,
  1991.

\bibitem{sch14}
Mac Schwager, Francesco Bullo, David Skelly, and Daniela Rus.
\newblock A ladybug exploration strategy for distributed adaptive coverage
  control.
\newblock In {\em 2008 IEEE International Conference on Robotics and
  Automation}, pages 2346--2353. IEEE, 2008.

\bibitem{sch4}
Mac Schwager, Daniela Rus, and Jean-Jacques Slotine.
\newblock Decentralized, adaptive coverage control for networked robots.
\newblock {\em The International Journal of Robotics Research}, 28(3):357--375,
  2009.

\bibitem{sch10}
Mac Schwager, Michael~P Vitus, Samantha Powers, Daniela Rus, and Claire~J
  Tomlin.
\newblock Robust adaptive coverage control for robotic sensor networks.
\newblock {\em IEEE Transactions on Control of Network Systems}, 4(3):462--476,
  2015.

\bibitem{ga2}
Rutuja Shivgan and Ziqian Dong.
\newblock Energy-efficient drone coverage path planning using genetic
  algorithm.
\newblock In {\em 2020 IEEE 21st International Conference on High Performance
  Switching and Routing (HPSR)}, pages 1--6. IEEE, 2020.

\bibitem{sch12}
Daniel~E Soltero, Mac Schwager, and Daniela Rus.
\newblock Decentralized path planning for coverage tasks using gradient descent
  adaptive control.
\newblock {\em The International Journal of Robotics Research}, 33(3):401--425,
  2014.

\bibitem{song2020coverage}
Cheng Song, Lu~Liu, Gang Feng, Yuan Fan, and Shengyuan Xu.
\newblock Coverage control for heterogeneous mobile sensor networks with
  bounded position measurement errors.
\newblock {\em Automatica}, 120:109118, 2020.

\bibitem{cassandras1}
Chuangchuang Sun, Shirantha Welikala, and Christos~G Cassandras.
\newblock Optimal composition of heterogeneous multi-agent teams for coverage
  problems with performance bound guarantees.
\newblock {\em Automatica}, 117:108961, 2020.

\bibitem{oc}
Qihai Sun, Ming Chi, Zhi-Wei Liu, and Dingxin He.
\newblock Observer-based coverage control of unicycle mobile robot network in
  dynamic environment.
\newblock {\em Journal of the Franklin Institute}, 2022.

\bibitem{ft}
Qihai Sun, Tianjun Liao, Zhi-Wei Liu, Ming Chi, and Dingxin He.
\newblock Fixed-time coverage control of mobile robot networks considering the
  time cost metric.
\newblock {\em Sensors}, 22(22):8938, 2022.

\bibitem{E_todescato2017multi}
Marco Todescato, Andrea Carron, Ruggero Carli, Gianluigi Pillonetto, and Luca
  Schenato.
\newblock Multi-robots gaussian estimation and coverage control: From
  client--server to peer-to-peer architectures.
\newblock {\em Automatica}, 80:284--294, 2017.

\bibitem{yadegar2021fault}
Meysam Yadegar and Nader Meskin.
\newblock Fault-tolerant control of nonlinear heterogeneous multi-agent
  systems.
\newblock {\em Automatica}, 127:109514, 2021.

\bibitem{yadegar2021mission}
Meysam Yadegar and Nader Meskin.
\newblock Mission independent fault-tolerant control of heterogeneous linear
  multiagent systems based on adaptive virtual actuator.
\newblock {\em International Journal of Adaptive Control and Signal
  Processing}, 35(3):401--419, 2021.

\bibitem{yadegar2021output}
Meysam Yadegar and Nader Meskin.
\newblock Output feedback fault-tolerant control of heterogeneous multi-agent
  systems.
\newblock {\em Asian Journal of Control}, 23(2):949--961, 2021.

\bibitem{global}
Yuxing Yang, Mao Su, Huijin Fan, Lei Liu, and Bo~Wang.
\newblock A constructive density function path leading to global coverage
  strategy for a gaussian random field.
\newblock In {\em 2023 IEEE 12th Data Driven Control and Learning Systems
  Conference (DDCLS)}, pages 249--254. IEEE, 2023.

\bibitem{zhang2021dirac}
Lin Zhang.
\newblock Dirac delta function of matrix argument.
\newblock {\em International Journal of Theoretical Physics}, 60(7):2445--2472,
  2021.

\bibitem{EB1}
Lei Zuo, Yang Shi, and Weisheng Yan.
\newblock Dynamic coverage control in a time-varying environment using bayesian
  prediction.
\newblock {\em IEEE transactions on cybernetics}, 49(1):354--362, 2017.

\bibitem{iet1}
Lei Zuo, Weisheng Yan, and Maode Yan.
\newblock Efficient coverage algorithm for mobile sensor network with unknown
  density function.
\newblock {\em IET Control Theory \& Applications}, 11(6):791--798, 2017.

\end{thebibliography}
	
\end{document}